\documentclass[aps,pra,onecolumn,notitlepage,10pt,superscriptaddress]{revtex4-2}
\usepackage{amsmath,amssymb,amsthm,mathtools}
\usepackage{graphicx,comment}
\usepackage{bm}
\usepackage{nccmath}
\usepackage{subcaption}
\usepackage{ragged2e}
\usepackage[colorlinks, citecolor=blue, linkcolor=red]{hyperref}
\usepackage{braket}
\newtheorem{theorem}{Theorem}

\usepackage{xcolor, color, soul}
\usepackage{multirow}

\begin{document}
	\title{
    %Non-Markovianity, P-divisibility and Information Back-flow
    Characterization of a damping channel as a mixture of amplitude damping and anti-damping channels of different parameters}
	\author{Vijay Pathak}
\email{vijaypathak.iisertvm@gmail.com}
\affiliation{Theoretical Physics Department, Poornaprajna Institute of Scientific Research, Bengaluru, India 562 164}
\affiliation{Anveshana, Swami Vivekananda Anusandhana Samsthana (S-VYASA), Bengaluru, Karnataka, India 560 105}

\author{R. Srikanth}
\email{srik@ppisr.res.in}
\affiliation{Theoretical Physics Department, Poornaprajna Institute of Scientific Research, Bengaluru, India 562 164}
	\date{\today}
	
	\begin{abstract}
Non-unital noise is a fundamental feature of open quantum systems, governing energy exchange and inducing state translations on the Bloch sphere. While such translations can be useful for tasks such as state preparation and channel capacity, unital dynamics where no translation occurs are often preferred in quantum information processing, particularly for error correction. Phase-covariant dynamics provide a general framework encompassing dissipation, excitation, and dephasing processes in qubit systems; however, commonly used models such as the generalized amplitude damping (GAD) channel offer only limited control over these features. In this work, we present a constructive framework for generating a broader class of phase-covariant dynamics by mixing amplitude-damping and anti-damping channels with unequal decay parameters and time-dependent mixing probabilities. This approach enables independent control over contraction and translation, allows continuous tuning between non-unital and unital regimes, and yields an effective dephasing contribution absent in GAD. To characterize these dynamics, we employ the general theorem for P-divisibility and show that it can be used constructively by evaluating it in suitably chosen Hilbert space bases, leading to simplified conditions for both unital and non-unital cases. The framework captures a wide range of behaviors, including CP-divisible, P-divisible but not CP-divisible, and non-P-divisible dynamics. We further show that appropriate mixing can reduce the deviation from the ideal noiseless evolution and that such improvement persists even after tuning the dynamics to the unital regime. Our results provide a flexible approach to engineering open system dynamics beyond standard thermal models, with potential applications in noise control and quantum information processing.
\end{abstract}
	
	\maketitle
	
\section{Introduction}
Noise is ubiquitous in quantum processes and plays a central role in determining the performance of quantum technologies. Amplitude-damping (AD) noise is an important example, with relevance to optical fiber communication \cite{gyongyosi2012secure}. AD noise \cite{nielsen2010quantum, preskill2015lecture}. AD is inherently non-unital and drives quantum states toward the ground state. Depending on the task, different features of open-system dynamics may instead be desirable: unital dynamics can be advantageous for quantum error correction, the translation associated with non-unital dynamics can be useful for channel capacity, and slower decay rates can improve the fidelity of quantum operations. Non-Markovian dynamics, characterized in part by information backflow, can also provide useful control over the evolution of quantum systems. Engineering open-system dynamics to achieve such desired properties therefore remains an important challenge. 

Amplitude damping is commonly associated with energy relaxation in a zero-temperature environment. At finite temperature, both excitation and relaxation processes become relevant, leading to the generalized amplitude damping (GAD) channel, which can be represented as a mixture of amplitude-damping and anti-amplitude-damping processes. In the GAD model, the system evolves toward a fixed point along the line connecting the ground and excited states on the Bloch sphere. However, the mixing probability and decay parameter are both determined by the properties of the system-environment interaction and the environment; hence, they cannot be controlled independently. More generally, convex mixing of quantum channels has been shown to be an effective method for generating diverse dynamical behaviors \cite{PhysRevA.101.062304, doi:10.1142/S1230161219500185, PhysRevA.103.042610, siudzinska2022phase}. Furthermore, such mixtures can be physically interpreted, for example, as arising from interaction with a squeezed thermal bath at non-zero temperature \cite{PhysRevA.77.012318}. 

AD, anti-AD, and GAD belong to the broader class of phase-covariant channels, which describe dissipation, excitation, and dephasing processes in qubit systems \cite{siudzinska2023geometry, siudzinska2023adjusting}. Various features of these dynamics, including non-monotonic population behavior and memory effects, have been explored in \cite{haase2019non, teittinen2018revealing}. Understanding and classifying such dynamics is closely tied to the notion of non-Markovianity. It has been shown \cite{pathak2026distinction} that information-theoretic measures do not capture all forms of non-Markovian dynamics, particularly in presence of non-unital evolution, whereas divisibility-based measures provide a more general characterization. In addition, Ref.~\cite{liu2013nonunital} introduces a criterion to quantify non-unitality, but it does not capture the information-flow aspect of the dynamics, while the role of mixing in the information-theoretic description of non-Markovianity has been discussed in \cite{breuer2018mixing}. Motivated by these complementary aspects, we analyze the dynamics using both divisibility- and BLP-based notions of non-Markovianity.

In this work, we consider a generalized mixing of amplitude-damping and anti-damping channels, in which the decay parameters of the two constituent channels can be different, and the mixing probability can be chosen independently of them. This extends the standard GAD scenario, where the decay parameters of the two constituent channels are identical. The resulting dynamics remain within the phase-covariant class but, unlike GAD, include an effective dephasing contribution. By appropriately choosing the decay parameters and mixing probability, we demonstrate that a wide range of dynamical behaviors can be realized, including transitions between non-unital and unital regimes. This tunability provides independent control over the contraction and translation of the Bloch sphere, enabling the engineering of distinct open-system dynamics for specific tasks, such as error mitigation, the realization of unital dynamics, and the control of information flow. The independent control of the mixing probability and decay parameters, absent in standard GAD construction, forms the basis of the present framework.

To characterize these dynamics, we employ the general theorem on P-divisibility and show that it can be used constructively by evaluating it in suitably chosen Hilbert-space bases, providing a simplified route to obtain conditions for both unital and non-unital channels. We further derive relations among the parameters of the phase-covariant dynamical map and use them to characterize different dynamical regimes through divisibility and distance-based measures. Finally, we study the deviation of the resulting channels from the ideal noiseless evolution and show that appropriate mixing can reduce this error. Notably, this improvement can persist even after tuning the dynamics to the unital regime, in contrast to the standard GAD case.

This paper is organized as follows. First, we present a method to determine P-divisibility conditions for non-unital channels using the general theorem. We then derive relations between the parameters of phase-covariant dynamical maps and characterize different dynamical regimes. Next, we implement these dynamics by mixing amplitude-damping and anti-damping channels, with both equal and unequal parameter values. Finally, we analyze the resulting error and discuss the implications of achieving unital dynamics through this framework.

\section{Conditions for different divisible dynamics}

An invertible, completely positive dynamics can be described by the following general master, which can be written in Lindblad form: 

\begin{equation}
\label{Eq.MasterEquation}    
\dfrac{d\rho}{dt}=-i[H(t),\rho(t)] + \sum_{j}\gamma_j(t)[A_j(t)\rho(t)A_j^{\dagger}(t) - \dfrac{1}{2}\{A_j^\dagger(t)A_j(t),\rho(t)\}]
\end{equation}
If the operators $A_j$ and rates $\gamma_j$ are time-dependent, the dynamics is not a semigroup. The following theorem gives the general condition for CP-divisibility and P-divisibility \cite{wissmann2015generalized}.
\begin{theorem}
    The dynamics generated by Eq.~\ref{Eq.MasterEquation} (I) is CP-divisible if and only if $\gamma_j(t)\geq0$ holds for all $j$ and $t\geq0$, and (II) is P-divisible if and only if for all $n\neq m$
    \begin{equation}
        \label{Eq.P-div}
        \sum_j\gamma_j(t)|\bra{m}A_j(t)\ket{n}|^2\geq0
    \end{equation}
    holds for any orthogonal basis $\ket{n}$ of $\mathcal{H}$ and all $t\geq0$.
    \label{thm:P-div}
\end{theorem}

While the condition in Eq.~(\ref{Eq.P-div}) is general, it is not straightforward to apply in practice, as it must hold for all orthonormal bases. Moreover, deriving positivity conditions directly from the dynamical map can be cumbersome, especially for non-unital dynamics, where the resulting constraints are often less transparent. Although direct approaches have been used in specific cases \cite{wissmann2015generalized}, they are not always easy to generalize. In what follows, we show that by choosing suitable bases adapted to the operator structure, the theorem can be used constructively to reduce the condition to a finite set of tractable constraints. This provides a practical method for determining P-divisibility for both unital and non-unital dynamics.

For unital dynamics with Lindblad operators $\sigma_i$ (Pauli operators), Theorem~\ref{thm:P-div} gives the necessary and sufficient condition for P-divisibility as the positivity of the pairwise rate sums, $\forall_{i\ne j} \gamma_i(t)+\gamma_j(t)\geq0$. To prove this, we consider three sets of basis vectors, corresponding to the two eigenvectors of each of the Pauli $\sigma$ operators $X$, $Y$, and $Z$, respectively.
\begin{comment}
\begin{itemize}
    \item X Eigenvectors: $\ket{n}\equiv\dfrac{1}{\sqrt{2}}\left(\begin{array}{c}
   1  \\
   1  \\
\end{array}\right)$, $\ket{m}\equiv\dfrac{1}{\sqrt{2}}\left(\begin{array}{c}
   1  \\
   -1  \\
\end{array}\right)$

    \item Y Eigenvectors: $\ket{n}\equiv\dfrac{1}{\sqrt{2}}\left(\begin{array}{c}
   1  \\
   i  \\
\end{array}\right)$, $\ket{m}\equiv\dfrac{1}{\sqrt{2}}\left(\begin{array}{c}
   1  \\
   -i  \\
\end{array}\right)$
    \item Z Eigenvectors: $\ket{n}\equiv\left(\begin{array}{c}
   1  \\
   0  \\
\end{array}\right)$, $\ket{m}\equiv\left(\begin{array}{c}
   0  \\
   1  \\
\end{array}\right)$
\end{itemize} 
\end{comment}
For the unital channel, Eq.~(\ref{Eq.P-div}) yields
\begin{equation*}
    \gamma_1(t)|\bra{m}X\ket{n}|^2 + \gamma_2(t)|\bra{m}Y\ket{n}|^2 + \gamma_3(t)|\bra{m}Z\ket{n}|^2\geq0.
\end{equation*}

For the $X$ basis, the condition becomes $\gamma_2(t)+\gamma_3(t)\geq0$; for the $Y$ basis, $\gamma_1(t)+\gamma_3(t)\geq0$; and for the $Z$ basis, $\gamma_1(t)+\gamma_2(t)\geq0$. Exchanging $\ket{n}$ and $\ket{m}$ does not alter these conditions. These three conditions are necessary by Theorem~\ref{thm:P-div}; we now show that they are also sufficient by verifying the inequality in an arbitrary basis.

Consider an arbitrary basis $\mathfrak{B} \equiv \left\{\ket{m}\equiv\left(\begin{array}{c}
   \alpha  \\
   \beta  \\
\end{array}\right), \ket{n}\equiv\left(\begin{array}{c}
   \beta^*  \\
   -\alpha^*  \\
\end{array}\right)\right\},$
where $|\alpha|^2 + |\beta|^2 =1$. 
The left-hand side of Eq.~(\ref{Eq.P-div}) then becomes 
$
    \gamma_1(t)|\beta^2-\alpha^2|^2 + \gamma_2(t)|\beta^2+\alpha^2|^2 + 4\gamma_3(t)|\beta|^2|\alpha|^2,
$
which can be rewritten as
\begin{equation*}
A(\gamma_1(t)+\gamma_2(t))+B(\gamma_1(t)+\gamma_3(t))+C(\gamma_2(t)+\gamma_3(t))
\end{equation*}
where
\begin{align*}
    A &=\dfrac{|\beta^2+\alpha^2|^2+|\beta^2-\alpha^2|^2-4|\beta|^2|\alpha|^2}{2}\\
B &=\dfrac{-|\beta^2+\alpha^2|^2+|\beta^2-\alpha^2|^2+4|\beta|^2|\alpha|^2}{2}\\
C &=\dfrac{|\beta^2+\alpha^2|^2-|\beta^2-\alpha^2|^2-4|\beta|^2|\alpha|^2}{2}.
\end{align*}

One can verify that $A$, $B$, and $C$ are non-negative for any $\alpha$ and $\beta$. For example, using the identity (valid for any complex $x, y$) that $|x+y|^2 + |x-y|^2=2(|x|^2+|y|^2)$ and setting $x=\beta^2, y=\alpha^2$, we find $A = (|\beta|^2-|\alpha|^2)^2$, which is always non-negative and vanishes if and only if $|\alpha|^2 = |\beta|^2 = \frac{1}{2}$. Similarly for $B, C$. Hence, the left-hand side of Eq.~(\ref{Eq.P-div}) for an arbitrary basis is always non-negative, provided the pairwise rate sums are non-negative.

For non-unital channels in the phase-covariant class, with Lindblad operators $\sigma_+\equiv\dfrac{1}{2}(X+iY)$, $\sigma_-\equiv\dfrac{1}{2}(X-iY)$ and $Z$, the necessary and sufficient condition for P-divisibility is \cite{filippov2020phase}:
\begin{equation}
   \gamma_\pm(t) \geq 0;~~ 2\gamma_z(t)+\sqrt{\gamma_+(t)\gamma_-(t)}\geq0
   \label{eq:Pdiv}
\end{equation}
To verify these conditions, we first consider the eigenvectors of $X$ and $Y$, which give
 \begin{equation*}
     \gamma_+(t) + \gamma_-(t) +4\gamma_z(t)\geq0.
 \end{equation*}
The $X$ and $Y$ eigenbases give the same condition, and exchanging the order of the eigenstates does not introduce an additional constraint. For the $Z$ eigenbasis, the two possible orderings of the eigenstates give the conditions
\begin{equation*}
     \gamma_+(t)\geq0,~ \gamma_-(t)\geq0.
\end{equation*}
For a general orthonormal basis $\mathfrak{B}$, the two possible orderings of the basis states give, from Eq.~(\ref{Eq.P-div}),
 \begin{align}
     \gamma_+(t)|\alpha|^4 + \gamma_-(t)|\beta|^4+4\gamma_z(t)|\alpha|^2|\beta|^2 &\geq0 \nonumber \\
     \gamma_+(t)|\beta|^4 + \gamma_-(t)|\alpha|^4+4\gamma_z(t)|\alpha|^2|\beta|^2 &\geq0.
 \end{align}
Using the necessary conditions obtained above, the first inequality can be rewritten as
 \begin{equation*}
     A\gamma_+(t) + B\gamma_-(t) + C(\gamma_+(t)+\gamma_-(t)+4\gamma_z(t))
 \end{equation*}
 where, 
 \begin{align*}
    A &=|\alpha|^2(|\alpha|^2-|\beta|^2)\\
B &=|\beta|^2(|\beta|^2-|\alpha|^2)\\
C &=|\alpha|^2|\beta|^2
\end{align*}

Since either $A$ or $B$ is always negative, the conditions $\gamma_+(t)>0$, $\gamma_-(t)>0$, and $\gamma_+(t)+\gamma_-(t)+4\gamma_z(t)>0$ are necessary but not sufficient for P-divisibility. However, if $2\gamma_z(t)\geq-\sqrt{\gamma_+(t)\gamma_-(t)}$ then
  \begin{equation*}
     \gamma_+(t)|\alpha|^4 + \gamma_-(t)|\beta|^4+4\gamma_z(t)|\alpha|^2|\beta|^2\geq(\gamma_+(t)|\alpha|^2-\gamma_-(t)|\beta|^2)^2\geq0
 \end{equation*}
This inequality holds for all $\alpha$, $\beta$, and $t$ provided $\gamma_+(t)$ and $\gamma_-(t)$ are positive. Hence $\sqrt{\gamma_+(t)\gamma_-(t)}+2\gamma_z(t)>0$ provides a sufficient condition for P-divisibility. These conditions will be used in the following sections to characterize the dynamics generated through channel mixing.

\section{Non-Markovian hierarchy for the phase-covariant dynamics}
Here, we determine the conditions on the parameters of the dynamical map that give rise to a hierarchy of dynamical behaviors, ranging from the least restrictive class of CPTP maps, through the distance-based notion of BLP-Markovianity and the divisibility-based notions of P-divisibility and CP-divisibility, to the most restrictive class of dynamical semigroups. All conditions are derived under the assumption that the dynamics from the initial time $ t=0$ to any final time $t$ is completely positive. The general master equation describing dissipation, excitation, and dephasing processes is  
\begin{equation}
    \dfrac{d\rho}{dt} = \gamma_+(t)(\sigma_{+}\rho\sigma_{-} - \dfrac{1}{2}\{\sigma_{-}\sigma_{+}, \rho\}) + \gamma_-(t)(\sigma_{-}\rho\sigma_{+} - \dfrac{1}{2}\{\sigma_{+}\sigma_{-}, \rho\}) +\gamma_z(t)(Z\rho Z-\rho)
\end{equation}
where, $\gamma_i(t)$ are the corresponding time-dependent rates. In the affine representation, the corresponding dynamical map can be written as
\begin{equation}\label{Eq: semigroup_map}
    \mathcal{E}(t) = \left(\begin{array}{cccc}
   1 & 0 & 0 & 0 \\
   0 & A(t) & 0 & 0 \\
   0 & 0 & A(t) & 0 \\
   X(t) & 0 & 0 &  B(t)
\end{array}\right)
\end{equation}
where the rates are related to the map parameters by
 \begin{equation}
     \begin{split}
         \gamma_\pm(t)&=\frac{1}{2}\bigg(-\frac{\dot{B}(t)}{B(t)}(1\pm X(t))\pm\dot{X}(t)\bigg), \\
         \gamma_z(t)&=\frac{1}{4}\bigg(-2\frac{\dot{A}(t)}{A(t)}+\frac{\dot{B}(t)}{B(t)}\bigg).
     \end{split}
     \label{Eq:rates}
 \end{equation}
The initial condition $\mathcal{E}(0) = \mathbb{I}_4$ implies $A(0)=1=B(0)$ and $X(0)=0$. With this formulation, we characterize a hierarchy of increasingly restrictive dynamics.

\paragraph{Complete Positivity condition.} 
The dynamics is CP for all times starting from the initial time $t_0=0$ if 
 \begin{equation}
 |B(t)| + |X(t)| \leq 1, \quad 4A^2(t) X^2(t) \leq(1+B(t))^2.
\end{equation}
These conditions are equivalent to the Fujiwara–Algoet inequalities specialized to the phase-covariant case~\cite{ruskai2002analysis}. 
They also imply that $\dot{A}(t)$ and $\dot{B}(t)$ cannot be positive at the initial time, as this would lead to unphysical states.

\paragraph{BLP Markovianity.} 
The map parameters can be expressed in terms of the integrated rates $\Gamma_i(t)=\int_0^t \gamma_i(\tau)d\tau$ as
\begin{equation}
    \begin{split}
     A(t) &=\exp{[-\frac{1}{2}(\Gamma_+(t)+\Gamma_-(t)+4\Gamma_z(t))]} \\
     B(t) &=\exp{[-\Gamma_+(t)-\Gamma_-(t)]}
     \\
    X(t) &=\exp[-\Gamma_+(t)-\Gamma_-(t)]\int_0^t(\gamma_+(\tau)-\gamma_-(\tau))\exp[\Gamma_+(\tau)+\Gamma_-(\tau)]d\tau
    \end{split}
    \label{eq:ABX}
\end{equation} 
The BLP distance between two states is 
\[D(t) = \sqrt{A^2(t)(\delta^2 x + \delta^2y) + B^2(t)\delta^2z}\]
with time derivative
\begin{equation}
    \dot{D}(t) = \dfrac{2A(t)\dot{A}(t)(\delta^2 x + \delta^2y) + 2B(t)\dot{B}(t)\delta^2z}{2D(t)},
    \label{eq:Dd}
\end{equation}
where the initial states are represented in Bloch-vector form as $\rho_i=(x_i, y_i, z_i)$, such that $\delta x = x_1-x_2$, $\delta y = y_1 - y_2$, and $\delta z = z_1- z_2$. 
From Eq.~(\ref{eq:Dd}), BLP-Markovianity requires 
\begin{equation}
    \dot{A}(t)\leq 0,~\dot{B}(t)\leq 0.
    \label{eq:dotD}
\end{equation}
These conditions are necessary and sufficient for $\dot{D}(t)\le0$ for all possible pairs of initial states.

From Eq.~(\ref{Eq:rates}), we have 
\begin{subequations}
\begin{align}
    \gamma_+(t)&+\gamma_-(t)=-\frac{\dot{B}(t)}{B(t)},\label{eq:g1g2} \\
    \gamma_+(t)&+\gamma_-(t)+4\gamma_z(t)=-2\frac{\dot{A}(t)}{A(t)}. \label{eq:g1g2gz}
\end{align}
\label{eq:gg}
\end{subequations}
This, in light of Eq. (\ref{eq:dotD}), implies:
\begin{equation}
    \gamma_+(t)+\gamma_-(t)+4\gamma_z(t)\geq0;~~\gamma_+(t)+\gamma_-(t)\geq0.
    \label{eq:BPLm-gamma}
\end{equation} 
These conditions are necessary and sufficient for the phase-covariant channel to be BLP-Markovian.

\paragraph{P-divisibility.}
Comparing the BLP-Markovianity conditions in Eq.~\ref{eq:BPLm-gamma} with the P-divisibility condition in Eq.~(\ref{eq:Pdiv}), we find that the P-divisibility is generally more restrictive than BLP-Markovianity. In particular,
\begin{subequations}
\begin{align}
    &\gamma_\pm(t) \ge 0 \implies \gamma_+(t) + \gamma_-(t) \ge 0, 
    \label{eq:a}\\
    &\gamma_+(t) + \gamma_-(t) \ge 2\sqrt{\gamma_+(t)\gamma_-(t)}~~ (\text{given}~\gamma_\pm(t)\ge0), \label{eq:b}
\end{align}
\end{subequations}
where Eq. (\ref{eq:b}) comes from the Arithmetic Mean-Geometric Mean (AM-GM) inequality. Since the converse of Eq.~(\ref{eq:a}) does not hold in general, P-divisibility implies BLP-Markovianity, whereas the converse does not hold. A particularly important special case is unital phase-covariant dynamics, for which ($\gamma_+(t) = \gamma_-(t)$). In this case, $2\sqrt{\gamma_+(t)\gamma_-(t)} = \gamma_+(t) + \gamma_-(t)$, and the P-divisibility condition in Eq.~(\ref{eq:Pdiv}) reduces to the BLP-Markovianity condition in Eq.~(\ref{eq:BPLm-gamma}). Hence, for unital phase-covariant dynamics, P-divisibility and BLP-Markovianity are equivalent.

\paragraph{CP-divisibility.} The necessary and sufficient condition for CP-divisibility is the positivity of all the Lindblad rates, i.e., the GKSL condition,
\begin{equation}
    \gamma_+(t) \ge 0,~ \gamma_-(t) \ge0, \gamma_z(t) \ge 0. 
\end{equation}
If, in addition, the rates are time-independent, the dynamics reduces to the semigroup limit.

Thus, the different notions of quantum Markovianity form the following nested hierarchy, ordered from the broadest to the most restrictive class:
\begin{equation}
\text{BLP-Markovian}
\supset
\text{P-divisible}
\supset
\text{CP-divisible}
\supset
\text{Semigroup}.
    \label{eq:hierarchy}
\end{equation}
Correspondingly, the notions of non-Markovianity form the reverse hierarchy, from the BLP-non-Markovianity as the strongest deviation to non-CP-divisibility as the weakest. Note that, as per earlier literature, an even weaker notion of non-Markovianity was given by any deviation from the semigroup condition, via a non-delta memory kernel leading to colored noise. 

Below, we discuss how these different levels of non-Markovianity arise from the map parameters. For our purposes, $A(t)$ and $B(t)$ are positive for all $t$. From Eq.~(\ref{eq:Dd}), BLP-Markovian dynamics requires, $\dot{A}(t)\leq 0, \quad \dot{B}(t)\leq 0.$ 
\begin{enumerate}
    \item Consider the case of $\dot{X}(t)>0$. If $\frac{|\dot{B}(t)|}{B(t)} (1+X(t)) > \dot{X}(t)$, then both $\gamma_+(t)>0$ and $\gamma_-(t)>0$. Further, from Eq.~(\ref{Eq:rates}), if $2\frac{\dot{A}(t)}{A(t)}\le\frac{\dot{B}(t)}{B(t)}$ then $\gamma_z(t)\ge0$, and the dynamics is CP-divisible, and hence P-divisible. On the other hand, if $2\frac{\dot{A}(t)}{A(t)}>\frac{\dot{B}(t)}{B(t)}$ then $\gamma_z(t)<0$, and the map is non-CP-divisible. In this case, according to Eq.~(\ref{eq:Pdiv}), the map is P-divisible if $2\gamma_z(t)+\sqrt{\gamma_+(t)\gamma_-(t)}\geq0$. If $\frac{|\dot{B}(t)|}{B(t)}(1+X(t))<\dot{X}(t),$ then $\gamma_+(t)\geq0$ but $\gamma_-(t)\leq0$. Therefore, irrespective of the relation between $\frac{\dot{A}(t)}{A(t)}$ and $\frac{\dot{B}(t)}{B(t)}$, the dynamics is non-P-divisible.

    \item If $\dot{X}(t)<0$. Since $\gamma_+(t)$ and $\gamma_-(t)$ have symmetric forms under $\dot{X}(t)\rightarrow-\dot{X}(t)$, the conditions obtained above remain unchanged, with the roles of $\gamma_+(t)$ and $\gamma_-(t)$ interchanged. This also means, in correspondence to the $\dot{X}(t)>0$ case above, that we can never have both $\gamma_+(t)$ and $\gamma_-(t)$ rates negative.
\end{enumerate} 

If the processes are BLP non-Markovian, then at least one of the $\dot{A}(t), \dot{B}(t)$ is positive. According to the hierarchy in Eq.~(\ref{eq:hierarchy}), BLP-non-Markovianity represents the most general form of the non-Markovianity considered here, and therefore the dynamics is also non-P-divisible wherever it is BLP-non-Markovian.
\begin{description}
    \item[$\dot{A}(t)>0, \dot{B}(t)<0$] From Eq.~(\ref{Eq:rates}), $\gamma_z(t)$ has to be negative. If $\dot{X}(t)$ is chosen from Eq.~(\ref{Eq:rates}) such that both $\gamma_+(t)$ and $\gamma_-(t)$ are positive, then, using the AM-GM inequality together with Eq.~(\ref{eq:g1g2gz}), we obtain $2\gamma_z(t)+\sqrt{\gamma_+(t)\gamma_-(t)} \le 2\gamma_z(t)+\gamma_+(t) + \gamma_-(t) = -2\frac{\dot{A}(t)}{A(t)}<0$. Hence, the process is non-P-divisible. If $\dot{X}(t)$ is chosen such that either $\gamma_+(t)$ or $\gamma_-(t)$ is negative, then the process is non-P-divisible directly from Eq.~(\ref{eq:Pdiv}), irrespective of the value of $\gamma_z(t)$. Furthermore, from Eq.~(\ref{eq:g1g2}), all three rates cannot be negative in this case, since $\dot{B}(t)<0$ requires $\gamma_+(t)+\gamma_-(t)>0$. 
    
    \item[$\dot{B}(t)>0, \dot{A}(t)<0$] By Eq~(\ref{Eq:rates}), $\gamma_z(t)$ has to be positive. At least one of $\gamma_+(t)$ or $\gamma_-(t)$ is then negative, depending on the sign of $\dot{X}(t)$. Hence, the dynamics is non-P-divisible, irrespective of the value of $\dot{A}(t)$. By choosing $\dot{X}(t)<\frac{\dot{B}(t)}{B(t)}(1+X(t))$ for $\dot{X}(t)>0$ and $|\dot{X}(t)|<\frac{\dot{B(t)}}{B(t)}(1-X(t))$ for $\dot{X}(t)<0$ both $\gamma_+(t)$ and $\gamma_-(t)$ can be made negative, unlike in the above case. Here too, all three rates cannot be negative, since $\gamma_z(t)\ge0$, as noted above.

    \item[$\dot{A}(t) \ge0, \dot{B}(t)\ge0$] From Eq.~(\ref{eq:g1g2}), it follows that $\gamma_+(t)$ and $\gamma_-(t)$ cannot both be positive simultaneously, which entails non-P-divisibility. Other possibilities are allowed depending on the relative values of $A(t)$, $B(t)$, $X(t)$ and their derivatives. In this case, however, all three rates can be negative.
    
    \item[Semi-group processes] The rates are constant and positive. Solving Eq.~(\ref{Eq:rates}) for the parameters of the dynamical map, we obtain that 
    $A(t)=e^{-\frac{1}{2}(4\gamma_z+\gamma_++\gamma_-)t}$, $B(t)=e^{-(\gamma_++\gamma_-)t}$, and $X(t)=\dfrac{\gamma_+-\gamma_-}{\gamma_++\gamma_-}(1-e^{-(\gamma_++\gamma_-)t})$. The parameters $A(t)$ and $B(t)$ decrease monotonically and ultimately approach zero. Their decay rates differ, and hence their asymptotic timescales also differ. The translation term $X(t)$ increases monotonically and approaches $\frac{\gamma_+-\gamma_-}{\gamma_++\gamma_-}$ at long times. Thus, the final state depends on the rates. 
\end{description}
\bigskip

\section{Implementation of damping channels}

The previous section established conditions on the parameters of the phase-covariant dynamical map for different dynamical regimes, based on divisibility and the BLP criterion. The next step is to realize such dynamics physically. The conventional approach is to start from a system-environment Hamiltonian and a specified environmental state and derive the corresponding reduced dynamics. However, for a prescribed time dependence of the decay rates, determining a Hamiltonian that reproduces the desired dynamics is, in general, a nontrivial inverse problem. An alternative approach is to construct the dynamics by convexly mixing known completely positive (CP) processes,
\begin{equation}
\mathcal{E}(t)=\sum_{i=1}^N p_i(t)\mathcal{E}_i(t),
\end{equation}
where $\mathcal{E}_i(t)$ are known CP processes and $p_i(t)$ are the corresponding mixing probabilities satisfying $\sum_i p_i(t)=1$. Since a convex combination of CP maps is again CP, this approach guarantees complete positivity without requiring explicit knowledge of the environment or the system-environment interaction. 

Among the possible elementary CP processes, AD and anti-AD channels are particularly suitable because their physical implementations are well established. Moreover, their convex mixture naturally yields a phase-covariant dynamical map, which is generally non-unital. Unlike the conventional generalized amplitude damping (GAD) channel, where the decay parameter and the mixing probability are determined by the system-environment interaction, we allow the two constituent channels to have different decay parameters while treating the mixing probability as an independent quantity. This additional freedom enables independent control over the contraction and translation of the Bloch sphere and, consequently, the realization of different dynamical regimes.

\begin{comment}
$\textbf{Condition for BLP Markovianity}$:
 The dynamical map for the above unital case is
 \begin{equation*}
    \mathcal{E}(t,0)= \left(\begin{array}{cccc}
   1 & 0 & 0 & 0  \\
   0 & e^{-\Gamma_2(t)-\Gamma_3(t)} & 0 & 0 \\
   0 & 0 & e^{-\Gamma_1(t)-\Gamma_3(t)} & 0 \\
   0 & 0 & 0 & e^{-\Gamma_1(t)-\Gamma_2(t)}
\end{array}\right)
 \end{equation*}
where, $\Gamma_i(t)=\int_0^t \gamma_i(\tau)d\tau$. From the map, we can see that if the P-divisibility condition is satisfied, then this is BLP-Markovian. From the results of our previous paper, we know that for unital cases, BLP and GBLP are the same. Additionally, the connection between P-divisibility and GBLP indicates that P-divisibility is equivalent to GBLP Markovian.

The dynamical map of the above non-unital case is
\begin{equation*}
    \mathcal{E}(t,0)= \left(\begin{array}{cccc}
   1 & 0 & 0 & 0  \\
   0 & e^{-a(t)} & 0 & 0 \\
   0 & 0 & e^{-a(t)} & 0 \\
   X(t) & 0 & 0 & e^{-b(t)}
\end{array}\right)
 \end{equation*}
 where, $X(t)=\exp[-\Gamma_1(t)-\Gamma_2(t)]\int_0^t(\gamma_1(\tau)-\gamma_2(\tau))\exp[\Gamma_1(\tau)+\Gamma_2(\tau)]d\tau$, $a(t)=\frac{1}{2}(\Gamma_1(t)+\Gamma_2(t)+4\Gamma_3(t))$, and $b(t)=\Gamma_1(t)+\Gamma_2(t)$ 

 From the form of the map, we can see that the necessary conditions of P-divisibility are also the conditions for BLP-Markovianity. Since necessary conditions do not guarantee P-divisibility, BLP-Markovian can be non-P-divisible, a point already discussed in a previous paper.
\end{comment}

Accordingly, we consider a mixture of the two processes ($N=2$), one in which only $\gamma_+(t)$ is non-zero and the other in which only $\gamma_-(t)$ is non-zero. If the rates are chosen as $-\dfrac{\dot{\eta}(t)}{\eta(t)}$, these reduce to the conventional AD and anti-AD channels, which drive the Bloch sphere towards the south ($\ket{1}$) and north ($\ket{0}$) poles, respectively. In the present construction, however, the two channels are allowed to have different decay parameters, denoted by $\eta_1(t)$ and $\eta_2(t)$. The resulting dynamical map is therefore

\begin{equation}\label{Eq: generalized_amplitude_damping_map}
    \mathcal{E}(t) = p(t)\left(\begin{array}{cccc}
   1 & 0 & 0 & 0 \\
   0 & \sqrt{\eta_1(t)} & 0 & 0 \\
   0 & 0 & \sqrt{\eta_1(t)} & 0 \\
   \eta_1(t)-1 & 0 & 0 & \eta_1(t)
\end{array}\right) + 
q(t)\left(\begin{array}{cccc}
   1 & 0 & 0 & 0 \\
   0 & \sqrt{\eta_2(t)} & 0 & 0 \\
   0 & 0 & \sqrt{\eta_2(t)} & 0 \\
   1-\eta_2(t) & 0 & 0 & \eta_2(t)
\end{array}\right)
\end{equation}
where, $p(t)+q(t) = 1$, with $p(t)$ denoting the mixing probability.

Since each constituent map is completely positive for $0<\eta_1(t),\eta_2(t)\leq1$, their convex combination is also completely positive. The corresponding Kraus operators are 

\begin{center}
\begin{equation}\label{Eq: amplitude_damping_kraus}
  \begin{split}  
 K_1 = \sqrt{p(t)}\left(\begin{array}{cc}
   \sqrt{\eta_1(t)} & 0  \\
   0 & 1  \\
\end{array}\right), 
K_2 = \sqrt{p(t)}\left(\begin{array}{cc}
    0 & 0  \\
    \sqrt{1-\eta_1(t)} & 0  \\,
\end{array}\right),  \\
K_3 = \sqrt{q(t)}\left(\begin{array}{cc}
   1 & 0  \\
   0 & \sqrt{\eta_2(t)}  \\
\end{array}\right),
K_4 = \sqrt{q(t)}\left(\begin{array}{cc}
    0 & \sqrt{1-\eta_2(t)}  \\
   0  & 0  \\
\end{array}\right)
\end{split}
\end{equation}
\end{center}

Expressing the convex mixture in matrix form, one obtains the phase-covariant dynamical map
\begin{equation}\label{Eq: mixed_generalized_amplitude_damping_map}
    \mathcal{E}(t) = \left(\begin{array}{cccc}
   1 & 0 & 0 & 0 \\
   0 & A(t) & 0 & 0 \\
   0 & 0 & A(t) & 0 \\
 X(t) & 0 & 0 & B(t)
\end{array}\right)
\end{equation}
where, 
\[A(t)\equiv p(t)\sqrt{\eta_1(t)}+q(t)\sqrt{\eta_2(t)},\]
\[B(t)\equiv p(t)\eta_1(t) + q(t)\eta_2(t),\] and 
\[X(t)\equiv p(t)\eta_1(t)-q(t)\eta_2(t)+q(t)-p(t).\] 
Here, $0<A(t), B(t)\leq1$ and $|X(t)|\leq1$. 

The resulting phase-covariant map is therefore determined by three independently tunable parameters: the mixing probability $p(t)$ and the decay parameters $\eta_1(t)$ and $\eta_2(t)$. This independent tunability provides the flexibility required to engineer the different dynamical regimes identified in the previous section. Before considering the mixed dynamics, we first examine the properties of the individual channels. 

There is only one nonzero rate in each individual channel. Therefore, if $\dot{\eta}(t) \leq 0$ for all time $t$, the dynamics is CP-divisible; otherwise, it is non-P-divisible. From the trace-distance measure, the dynamics is BLP non-Markovian if $\dot{\eta}(t) > 0$ for some time $t$ and BLP Markovian otherwise. Thus, for these individual channels, divisibility-based Markovianity is equivalent to  BLP-Markovianity. 

Another important feature of the individual channels is that the translation term is not independent of the decay term. If the Bloch sphere shrinks, its translation decreases with time, whereas if the sphere temporarily expands, its translation also increases. Thus, Markovian dynamics in the individual channels can exhibit only monotonic translation. Furthermore, since only one rate is non-zero, no process can be P-divisible without also being CP-divisible. These individual channels form semigroups only when $\eta_i(t)=\exp{[-\gamma_it]}$. Finally, they cannot be unital, as this would require the corresponding decay rate to vanish.  

\subsection{Mixing of amplitude damping and anti-damping channels with equal decay coefficients $\eta_1(t)=\eta_2(t)$: Generalized amplitude damping channel}

We first consider the special case in which the two constituent channels have identical decay parameters, $\eta_1(t)=\eta_2(t)=\eta(t)$. The resulting dynamical map has the same form as the generalized amplitude damping (GAD) channel. However, unlike the conventional thermal derivation of GAD, where the mixing probability is fixed by the system-environment interaction (or bath temperature), here the mixing probability remains an independent parameter owing to the convex mixing construction. The resulting dynamical map is

\begin{equation}\label{Eq: generalized_amplitude_damping_map}
    \mathcal{E}(t) = \left(\begin{array}{cccc}
   1 & 0 & 0 & 0 \\
   0 & \sqrt{\eta(t)} & 0 & 0 \\
   0 & 0 & \sqrt{\eta(t)} & 0 \\
   (2p(t)-1)(\eta(t)-1) & 0 & 0 & \eta(t)
\end{array}\right) 
\end{equation}

Although each constituent channel has only one non-zero rate, their convex mixture is characterized by two non-zero rates, 

\[\gamma_+(t) = (\eta(t)-1)\dot{p}(t) - \dfrac{\dot{\eta}(t)q(t)}{\eta(t)},\] and 
\[\gamma_-(t) = (\eta(t)-1)\dot{q}(t) - \dfrac{\dot{\eta}(t)p(t)}{\eta(t)}.\]

Since the contraction parameters $A(t)$ and $B(t)$ depend only on the decay parameter $\eta(t)$, the BLP Markovianity of the dynamics is determined entirely by $\eta(t)$ and is independent of the mixing probability $p(t)$. In contrast, the translation parameter $X(t)$ depends explicitly on $p(t)$ and can therefore be varied without affecting the trace distance. Consequently, the mixing probability can alter the divisibility properties of the dynamics while leaving the BLP-Markovianity unchanged. Thus, this convex-mixing construction allows non-P-divisible yet BLP Markovian dynamics. Furthermore, since $\gamma_z(t)=0$, P-divisibility is equivalent to  CP-divisibility for this channel. 

\textbf{Condition 1.} Suppose $\dot{\eta}(t)\leq0$ for all time. The dynamics is then BLP-Markovian. Since $\dot{q}(t)=-\dot{p}(t)$, exactly one of $\dot{p}(t)$ and $\dot{q}(t)$ is positive at any instant. Consequently, at least one of the rates, $\gamma_+(t)$ or $\gamma_-(t)$, remains positive throughout the evolution. 
\begin{itemize}
    \item Whenever $\dot{p}(t)<0$ (equivalently, $\dot{q}(t)>0$), $\gamma_+(t)$ is positive. If
    \[\frac{|\dot{p}(t)|}{p(t)}<\frac{|\dot{\eta}(t)|}{\eta(t)(1-\eta(t))},\] then $\gamma_-(t)$ is also positive; otherwise, $\gamma_-(t)$ is negative.

   \item Whenever $\dot{p}(t)>0$ (equivalently, $\dot{q}(t)<0$), the corresponding conditions remain valid after interchanging $p(t)\leftrightarrow q(t)$ and $\gamma_+(t)\leftrightarrow\gamma_-(t)$.
\end{itemize}

\textbf{Global condition:}
The dynamics is CP-divisible if the corresponding positivity condition is satisfied whenever the relevant case applies, so that both $\gamma_+(t)$ and $\gamma_-(t)$ remain positive for all time. Otherwise, the dynamics is non-P-divisible.

\begin{figure}
\centering
    \begin{subfigure}{0.45\textwidth}
    \includegraphics[scale=0.5]{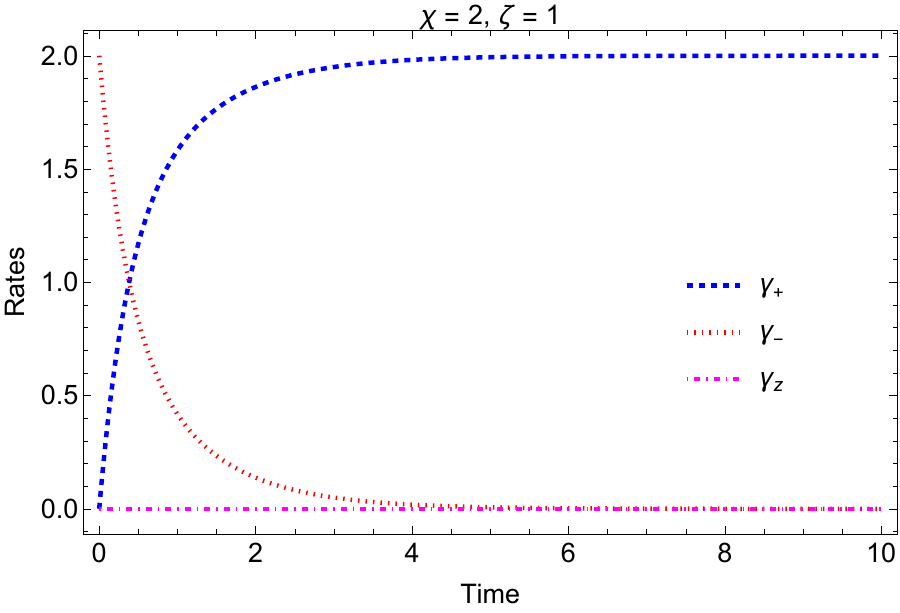} 
    \end{subfigure}
    %\hspace{0.1pt}
    \begin{subfigure}{0.45\textwidth}
    \includegraphics[scale=0.5]{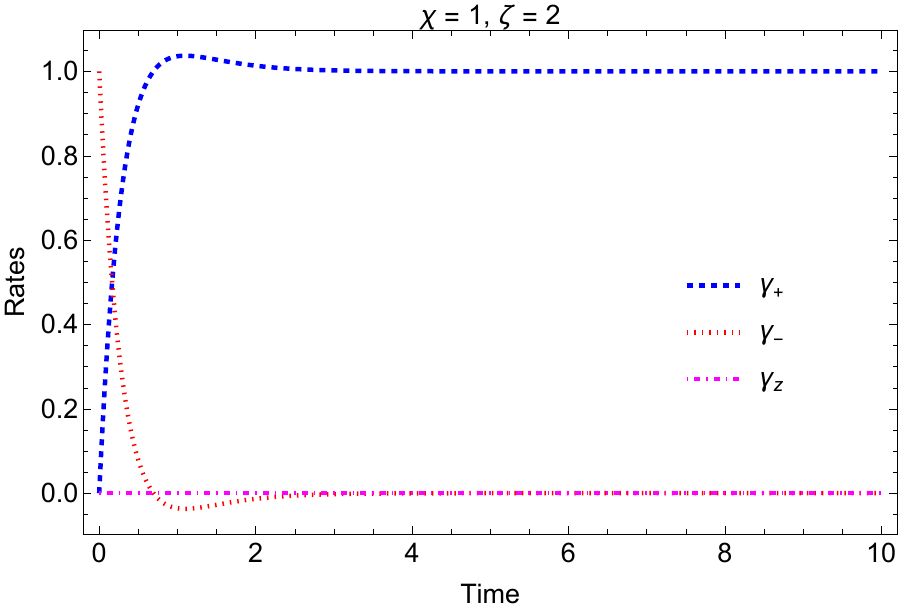}
    \end{subfigure}
\caption{\justifying
The canonical rates $\gamma_+(t)$, $\gamma_-(t)$, and $\gamma_z(t)$ are plotted for an exponentially decaying channel parameter $\eta(t)=\exp[-\chi t]$ and an exponentially decaying mixing probability
$p(t)=\exp[-\zeta t]$. \textbf{Left:} $\chi=2$ and $\zeta=1$, for which the resulting dynamics is CP-divisible. \textbf{Right:} $\chi=1$ and $\zeta=2$, for which the resulting dynamics is non-CP-divisible.}
\label{Decay_GAD}
\end{figure}

\textbf{Examples}:
    
(1) Let $\eta(t)=\exp{[-\chi t]}$ and $p(t)=\exp{[-\zeta t]}$. If $\chi>\zeta$, the corresponding positivity condition is satisfied for all time, so both rates remain positive, and the dynamics is CP-divisible (Left figure of Fig.~\ref{Decay_GAD}). If the condition is violated during part of the evolution, one of the rates becomes negative, and the dynamics becomes non-P-divisible (Right figure of Fig.~\ref{Decay_GAD}).

\begin{figure}
    \centering
    \begin{subfigure}{0.45\textwidth}
        \includegraphics[scale=0.48]{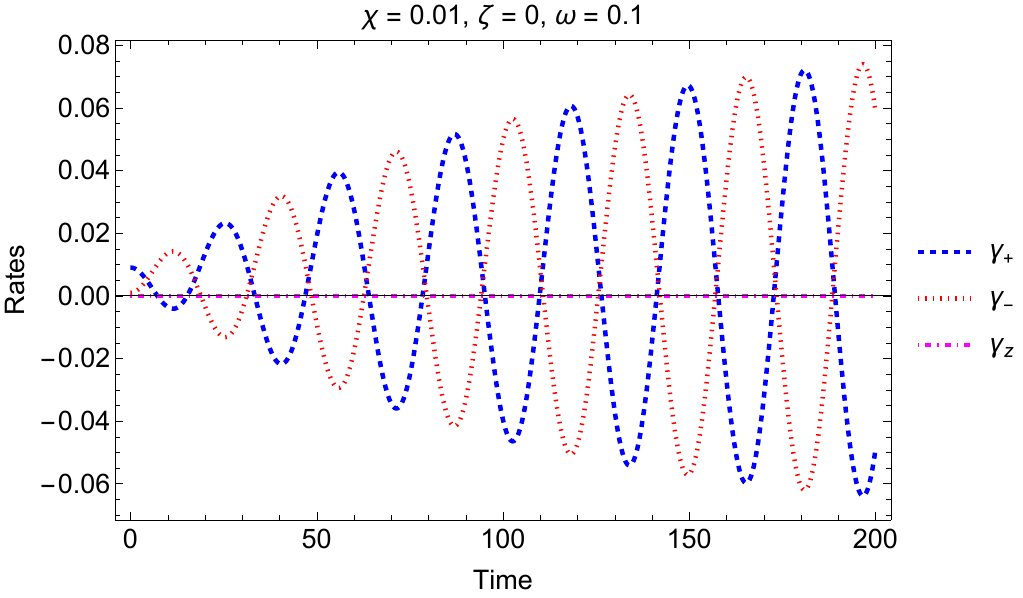} 
    \end{subfigure}
    \hspace{0.15cm}
    \begin{subfigure}{0.45\textwidth}
        \includegraphics[scale=0.48]{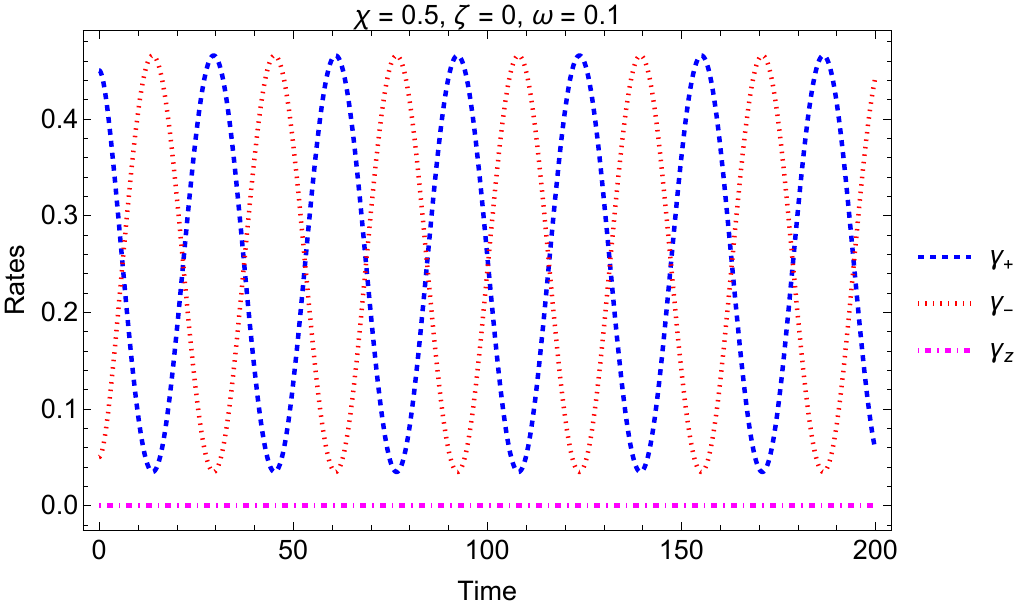} 
    \end{subfigure}\\
    \begin{subfigure}{0.45\textwidth}
        \includegraphics[scale=0.48]{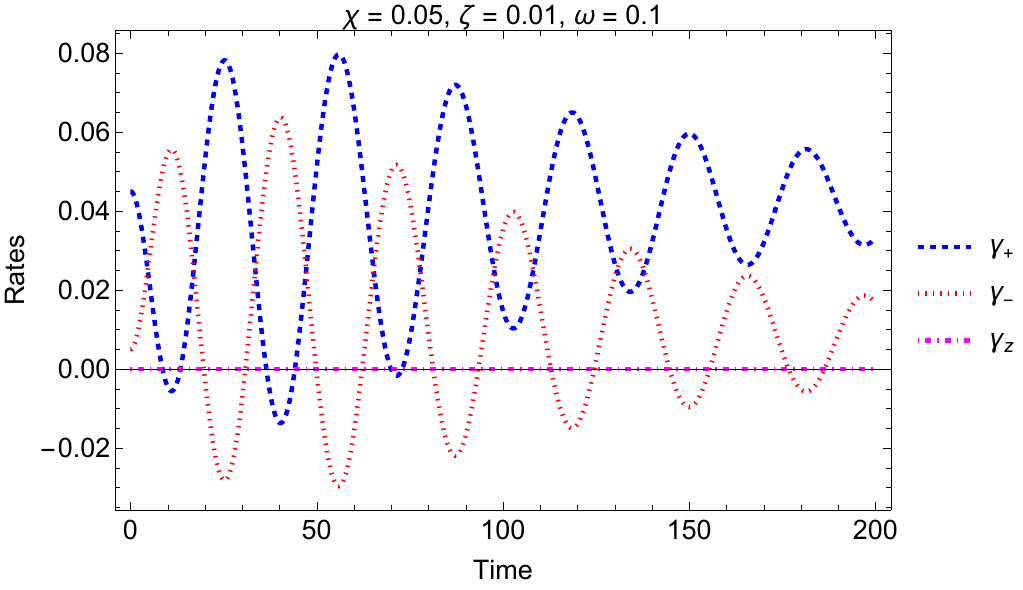}
    \end{subfigure}
    \hspace{0.15cm}
    \begin{subfigure}{0.45\textwidth}
        \includegraphics[scale=0.48]{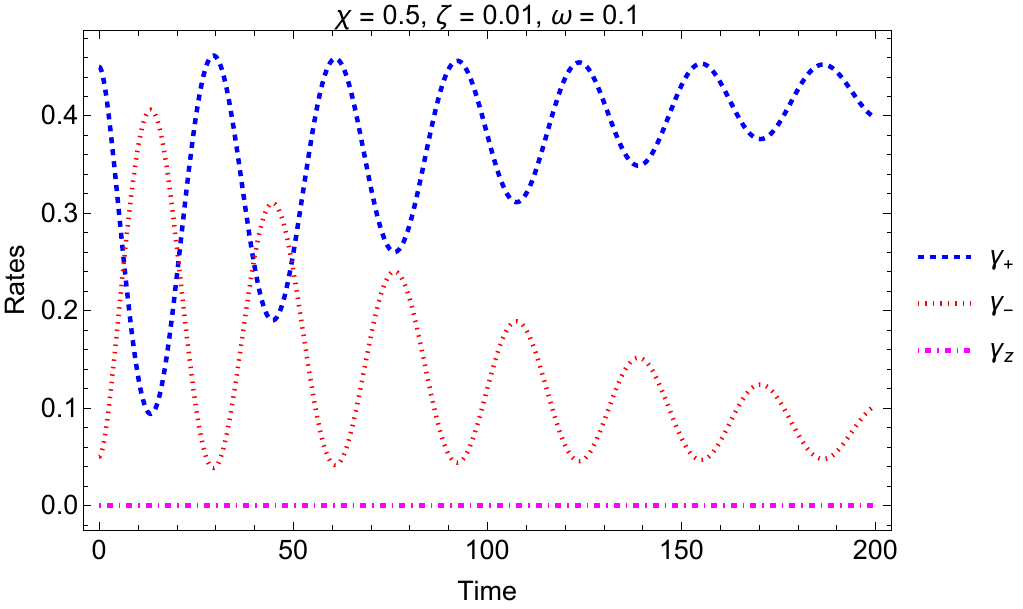}
    \end{subfigure}
    
    \caption{\justifying
The canonical rates $\gamma_+(t)$, $\gamma_-(t)$, and $\gamma_z(t)$ are plotted for the monotonically decaying channel parameter $\eta(t)=\exp[-\chi t]$ and the mixing probability $p(t)=p_0+p\exp[-\zeta t]\sin^2(\omega t)$. For all panels, $p_0=0.1$ and $p=0.8$. \textbf{Top left:} non-P-divisible dynamics for a purely oscillatory mixing probability ($\zeta=0$) with $\chi$ close to $\zeta$. \textbf{Top right:} CP-divisible dynamics for $\zeta=0$ with $\chi>\zeta$. \textbf{Bottom left:} non-P-divisible dynamics for an oscillator decaying mixing probability with $\zeta=0.01$. \textbf{Bottom right:} CP-divisible dynamics for the same decaying
mixing probability with a sufficiently large $\chi$.
The oscillatory behavior of the rates originates from the time dependence of the mixing probability, while $\gamma_z(t)=0$ for the conventional GAD channel.}
    \label{OscDecay_GAD}
\end{figure}

(2) The mixing probability need not be monotonic. It can oscillate with or without a decaying envelope while the decay parameter $\eta(t)$ remains monotonically decreasing. For example, consider \[p(t)=p_0+p\exp{[-\zeta t]\sin^2{\omega t}},\] whose effect on the rates is shown in Fig.~\ref{OscDecay_GAD}. Different forms of oscillatory mixing probabilities can therefore be realized while maintaining BLP-Markovianity. The oscillatory behavior of the rates arises entirely from the translation term, due to the time dependence of $p(t)$.

\textbf{Conclusion:}
Since
\[\gamma_+(t)+\gamma_-(t)=-\frac{\dot{\eta}(t)}{\eta(t)},\]
both rates cannot be made negative simultaneously. However, by suitably choosing the mixing probability, one of the rates can be made negative while the other remains positive, thereby realizing non-P-divisible yet BLP-Markovian dynamics. On the other hand, by choosing the mixing probability such that both rates remain positive throughout the evolution, the dynamics becomes CP-divisible. In particular, a constant mixing probability cannot realize any dynamics other than CP-divisible, irrespective of its value or the specific monotonic decay of $\eta(t)$.

\textbf{Condition 2.} If $\dot{\eta}(t)$ becomes positive during some intervals, the dynamics is BLP non-Markovian and, according to the P-divisibility theorem, must also be non-P-divisible. This is readily confirmed from the rate equations, which show that at least one of the rates $\gamma_+(t)$ or $\gamma_-(t)$ becomes negative during those intervals, irrespective of the behavior of the mixing probability $p(t)$. The sign of the other rate depends on the relative magnitudes of $\dot{p}(t)$ and $\dot{\eta}(t)$. 
\begin{itemize}
    \item[A] Whenever $\dot{\eta}(t)>0$,
    \begin{itemize}
        \item[1.] If $\dot{p}(t)<0$ (equivalently $\dot{q}(t)>0$), then $\gamma_-(t)$ is negative. If 
        \[\dfrac{\dot{q}(t)}{q(t)}>\dfrac{\dot{\eta}(t)}{\eta(t)(1-\eta(t))},\] then  $\gamma_+(t)$ is positive; otherwise $\gamma_+(t)$ is negative. 

        \item[2.] If $\dot{p}(t)>0$ (equivalently $\dot{q}(t)<0$), then $\gamma_+(t)$ is negative. If \[\dfrac{\dot{p}(t)}{p(t)}>\dfrac{\dot{\eta}(t)}{\eta(t)(1-\eta(t))},\] then  $\gamma_-(t)$ is positive; otherwise $\gamma_-(t)$ is negative. 
     \end{itemize}
    \item[B] During the intervals for which $\dot{\eta}(t)<0$, the dynamics should be analyzed using the conditions obtained in Condition 1.
\end{itemize}
 
\textbf{Conclusion:} Since \[ \gamma_+(t)+\gamma_-(t)=-\frac{\dot{\eta}(t)}{\eta(t)}, \] both rates can be made negative simultaneously. By suitably choosing the mixing probability, one can realize either one rate negative or both rates negative dynamics during the intervals in which $\dot{\eta}(t)>0$, making the dynamics non-P-divisible. In particular, for a constant mixing probability, both rates always have the same sign, determined solely by the sign of $\dot{\eta}(t)$. Thus, unlike Condition 1, even a constant mixing probability can realize BLP non-Markovian dynamics.

Finally, the semigroup dynamics is recovered when
\[\eta(t)=\exp[-(\gamma_++\gamma_-)t]\]
with constant rates and the mixing probability
\[p=\frac{\gamma_-}{\gamma_++\gamma_-}.\]
This semigroup dynamics is unital if and only if the two rates are equal, i.e., $\gamma_+=\gamma_-$. More generally, for arbitrary time-dependent dynamics, the resulting channel is unital if and only if $\gamma_+(t)=\gamma_-(t)$, which is equivalent to $p(t)=\frac{1}{2}$.

\subsection{Mixing of amplitude damping and anti-damping channels with unequal decay coefficients $\eta_1(t)\neq\eta_2(t)$}

Unlike the equal-decay case, the resulting dynamical map is, in general, not of the generalized amplitude form, although it remains phase-covariant. The independent decay parameters $\eta_1(t)$ and $\eta_2(t)$ together with the mixing probability $p(t)$ provide three independent controls over the dynamics. Consequently, a much richer set of dynamical behaviors can be realized than in the equal-decay case. The derivatives of $A(t)$, $B(t)$, and $X(t)$, which govern the resulting dynamics through the canonical rates, are given by 
\begin{equation}\label{generalAB}
    \begin{split}
        \dot{A}(t)&=p(t)\frac{\dot{\eta}_1(t)}{2\sqrt{\eta_1(t)}}+q(t)\frac{\dot{\eta}_2(t)}{2\sqrt{\eta_2(t)}}+\dot{p}(t)(\sqrt{\eta_1(t)}-\sqrt{\eta_2(t)}) \\ 
        \dot{B}(t)&=p(t)\dot{\eta}_1(t)+q(t)\dot{\eta}_2(t)+\dot{p}(t)(\eta_1(t)-\eta_2(t)) \\
        \dot{X}(t)&=p(t)\dot{\eta}_1(t)-q(t)\dot{\eta}_2(t)+\dot{p}(t)(\eta_1(t)+\eta_2(t)-2) 
    \end{split}
\end{equation}

These expressions lead to the following form of the rates
\begin{equation}\label{Generalrates}
    \begin{split}
        \gamma_+(t)&=\frac{-q(t)[\dot{\eta}_2(t)(q(t)+p(t)\eta_1(t)) + p(t)\dot{\eta}_1(t)(1-\eta_2(t))]-\dot{p}(t)\eta_1(t)(1-\eta_2(t))}{p(t)\eta_1(t)+q(t)\eta_2(t)} \\ 
        \gamma_-(t)&=\frac{-p(t)[\dot{\eta}_1(t)(p(t)+q(t)\eta_2(t)) + q(t)\dot{\eta}_2(t)(1-\eta_1(t))]+\dot{p}(t)\eta_2(t)(1-\eta_1(t))}{p(t)\eta_1(t)+q(t)\eta_2(t)} \\ 
        \gamma_z(t)&=\frac{\frac{p(t)q(t)}{\sqrt{\eta_1(t)\eta_2(t)}}d_1(t)(\dot{\eta}_1(t)\eta_2(t)-\dot{\eta}_2(t)\eta_1(t))+\dot{p}(t)d_1(t)^2(\sqrt{\eta_2(t)}+p(t)d_2(t))}{4(p(t)\sqrt{\eta_1(t)}+q(t)\sqrt{\eta_2(t)})(p(t)\eta_1(t)+q(t)\eta_2(t))}
    \end{split}
\end{equation}
where $d_1(t)\equiv\sqrt{\eta_1(t)}-\sqrt{\eta_2(t)}$ and $d_2(t)\equiv\sqrt{\eta_1(t)}+\sqrt{\eta_2(t)}$. 

For the special case of a constant mixing probability, i.e., $\dot{p}(t)=0$, these expressions reduce to 

\begin{equation}\label{ConstantPAB}
    \begin{split}
        \dot{A}(t)&=p\frac{\dot{\eta}_1(t)}{2\sqrt{\eta_1(t)}} + q\frac{\dot{\eta}_2(t)}{2\sqrt{\eta_2(t)}} \\
        \dot{B}(t)&=p\dot{\eta}_1(t) + q\dot{\eta}_2(t) \\
        \dot{X}(t)&=p\dot{\eta}_1(t) -q\dot{\eta}_2(t)
    \end{split}
\end{equation} 
with the corresponding canonical rates given by

\begin{equation}\label{rateConstantP}
    \begin{split}
        \gamma_+(t) &=\frac{-q[\dot{\eta}_2(t)(q+p\eta_1(t)) + p\dot{\eta}_1(t)(1-\eta_2(t))]}{p\eta_1(t)+q\eta_2(t)} \\
        \gamma_-(t) &=\frac{-p[\dot{\eta}_1(t)(p+q\eta_2(t)) + q\dot{\eta}_2(t)(1-\eta_1(t))]}{p\eta_1(t)+q\eta_2(t)} \\
        \gamma_z(t) &=\frac{pq(\sqrt{\eta_1(t)}-\sqrt{\eta_2(t)})(\dot{\eta}_1(t)\eta_2(t)-\dot{\eta}_2(t)\eta_1(t))}{4\sqrt{\eta_1(t)\eta_2(t)}(p\sqrt{\eta_1(t)}+q\sqrt{\eta_2(t)})(p\eta_1(t)+q\eta_2(t))}
    \end{split}
\end{equation}

The characterization naturally separates into the following three classes according to the properties of the constituent channels:
\begin{itemize}
    \item \textbf{Condition 1:} Both constituent channels are CP-divisible.
    \item \textbf{Condition 2:} Both constituent channels are non-P-divisible (equivalently BLP non-Markovian).
    \item \textbf{Condition 3:} One constituent channel is CP-divisible while the other is non-P-divisible.
\end{itemize}
Within each condition, the resulting dynamics is analyzed separately for constant and time-dependent mixing probabilities.

\textbf{Condition 1.} Both constituent channels are CP-divisible, i.e., 
\[\dot{\eta}_1(t)\leq0. \qquad \dot{\eta}_2(t)\leq0 \] 
for all times. We now analyze the resulting dynamics for different choices of the mixing probability. 

\textbf{\textit{Case a.}} We first consider a constant mixing probability, i.e., $\dot{p}(t)=0$. Using Eqs.~(\ref{ConstantPAB}) and (\ref{rateConstantP}), the resulting dynamics is characterized as follows.

Since both decay parameters decrease monotonically, $\dot{A}(t)$ and $\dot{B}(t)$ are also negative for all time. Consequently, the trace distance decreases monotonically, implying that the resulting dynamics is BLP-Markovian. The same conditions also guarantee that both $\gamma_+(t)$ and $\gamma_-(t)$ remain positive throughout the evolution. The sign of $\gamma_z(t)$. however, depends on the relative behavior of $\eta_1(t)$ and $\eta_2(t)$ and can therefore be either positive or negative. Consequently, the dynamics may be CP-divisible, P-divisible, or non-P-divisible while remaining BLP Markovian.  

\textbf{Example 1.} If $\eta_1=e^{-\chi_1 t}$ and $\eta_2=e^{-\chi_2 t}$ then the numerator of $\gamma_z(t)$ will be $pq\eta_1(t)\eta_2(t)(\chi_2-\chi_1)(\sqrt{\eta_1(t)}-\sqrt{\eta_2(t)})$. Since the numerator remains positive for all time, $\gamma_z(t)$ is also positive. Therefore, all three rates remain positive, and the resulting dynamics is CP-divisible. 

\begin{figure}[t]
    \centering
    \begin{subfigure}{0.45\textwidth}
        \includegraphics[scale=0.5]{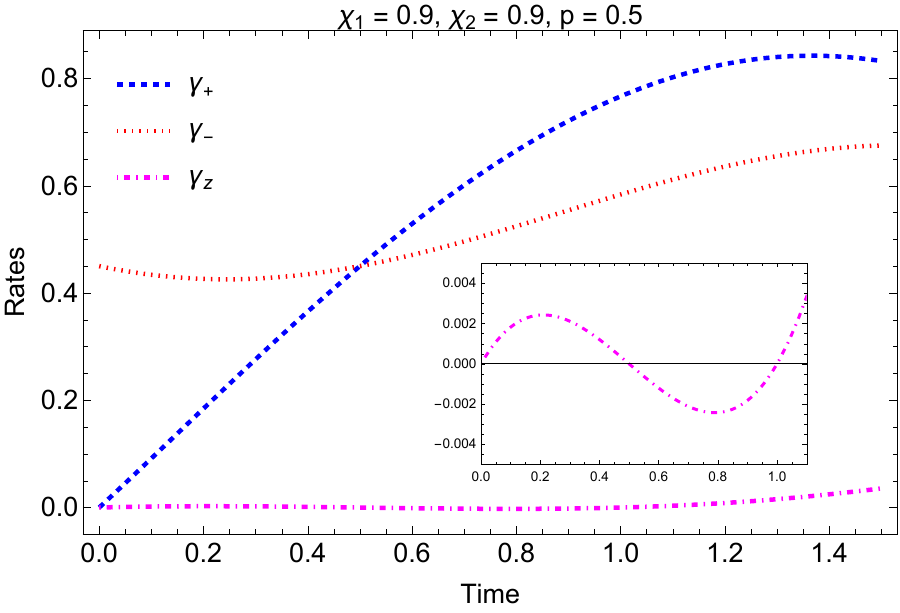}
    \end{subfigure}
    \begin{subfigure}{0.45\textwidth}
        \includegraphics[scale=0.5]{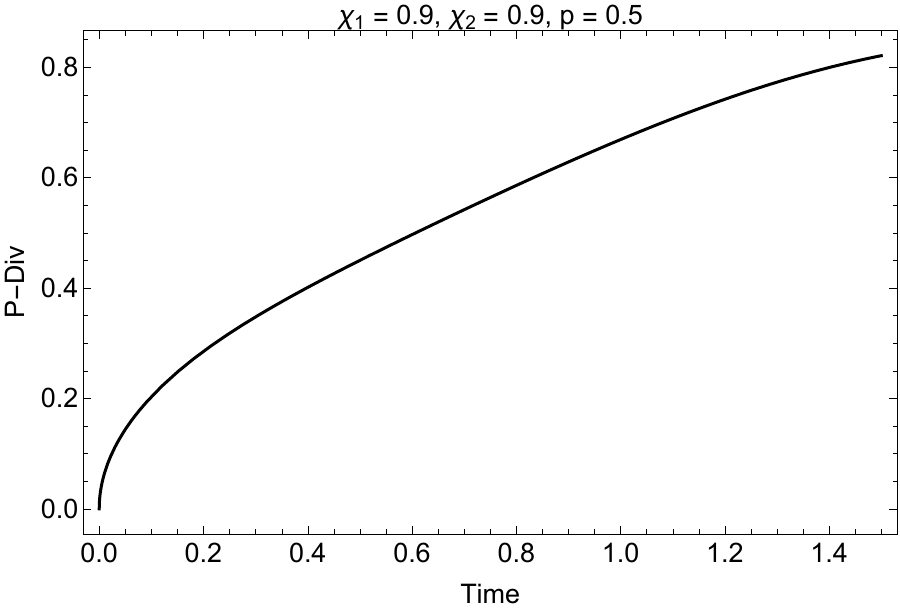}
    \end{subfigure}
    \caption{\justifying
\textbf{Left:} The canonical rates $\gamma_+(t)$, $\gamma_-(t)$, and $\gamma_z(t)$ are plotted for
$\eta_1(t)=\exp[-\chi_1 t]$ and $\eta_2(t)=\exp[-\chi_2 t^2]$ with constant mixing probability $p$.
The inset shows the region where $\gamma_z(t)<0$.
\textbf{Right:} The P-divisibility condition $2\gamma_z(t)+\sqrt{\gamma_+(t)\gamma_-(t)}$ is plotted and remains positive throughout the evolution, demonstrating P-divisible but non-CP-divisible dynamics.}
    \label{Pdiv}
\end{figure}

\begin{figure}[t]
\centering
    \includegraphics[scale=0.75]{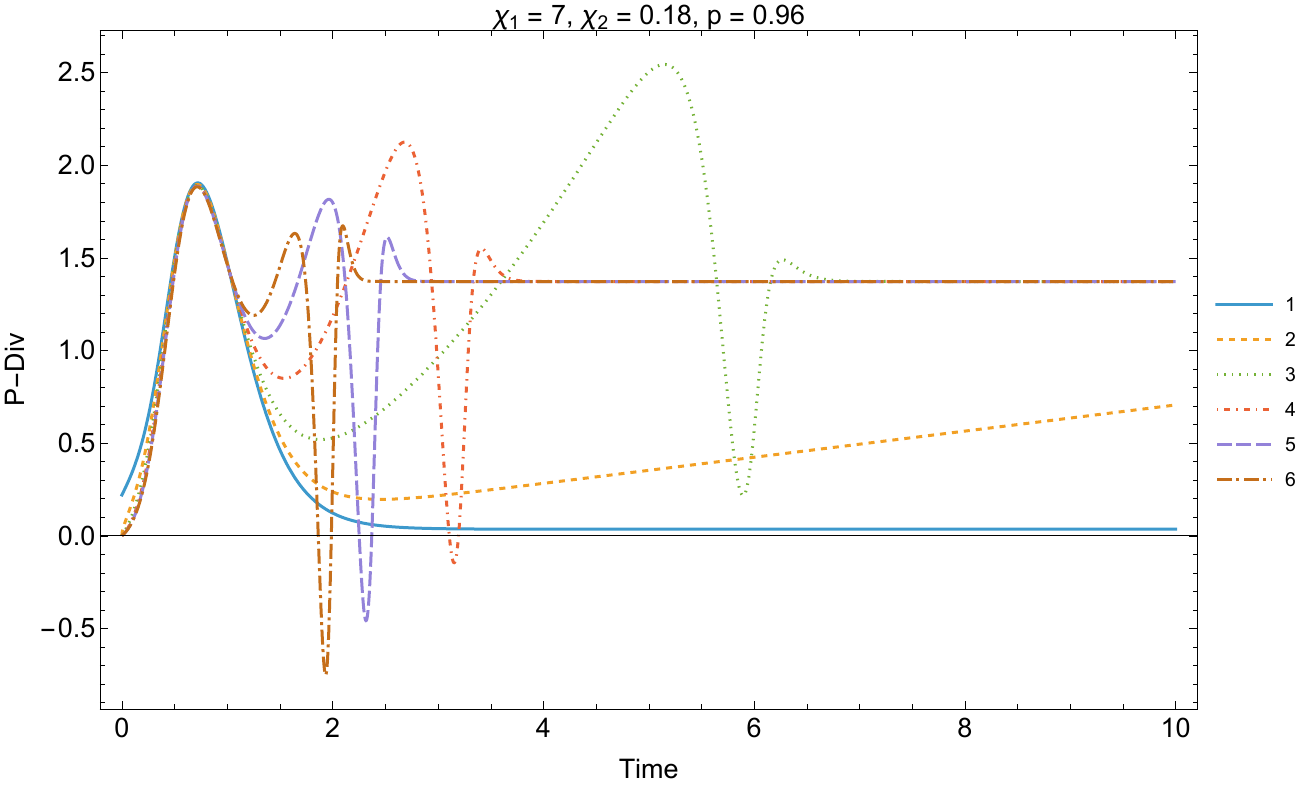}
        \caption{\justifying
The P-divisibility condition is plotted as a function of time for $\eta_1(t)=\exp[-\chi_1 t]$ and
$\eta_2(t)=\exp[-\chi_2 t^i]$ with constant mixing probability $p$. The different curves correspond to different values of the exponent $i$. The condition remains positive throughout the evolution for
$i\leq 3$, indicating P-divisible dynamics, whereas for $i>3$ it becomes negative over part of the evolution, indicating non-P-divisible dynamics.}
    \label{nonPdiv}
\end{figure}

\textbf{Example 2.} If one of the decay parameters is chosen to be non-exponential, the rate $\gamma_z(t)$ can become negative at times even though both constituent channels remain CP-divisible. For example, let
\[\eta_1(t)=e^{-\chi_1 t} \qquad \eta_2(t)=e^{-\chi_2 t^2}.\] 
With an appropriate choice of $\chi_1$ and $\chi_2$, $\gamma_z(t)$ becomes negative during part of the evolution, as shown in the inset of the left panel of Fig.~\ref{Pdiv}. Nevertheless, 
\[2\gamma_z(t)+\sqrt{\gamma_+(t)\gamma_-(t)}>0,\]
so the resulting dynamics remains P-divisible despite not being CP-divisible, as shown in the right panel of Fig.~\ref{Pdiv}. Furthermore, replacing 
\[\eta_2(t)=e^{-\chi_2 t^i}\]
allows a transition from P-divisible to non-P-divisible dynamics. The exact transition point depends on the remaining parameters. One such transition, occurring at $i=3$, is shown in Fig.~\ref{nonPdiv}.

\textbf{\textit{Conclusion:}}
With a constant mixing probability, the resulting dynamics is always BLP Markovian. By appropriately choosing the decay parameters, one can realize CP-divisible, P-divisible, and non-P-divisible dynamics without making either constituent channel non-Markovian. Throughout the evolution, both $\gamma_+(t)$ and $\gamma_-(t)$ remain positive, while the sign of $\gamma_z(t)$ determines the divisibility class. Consequently, BLP non-Markovian dynamics cannot be realized in this case.

\textbf{\textit{Case b.}} We now allow the mixing probability to decrease monotonically, i.e., $\dot{p}(t)\leq0$ for all time. Using Eqs.~(\ref{generalAB}) and (\ref{Generalrates}), the resulting dynamics is characterized as follows.

Since $\dot{p}(t)\leq0$, $\dot{\eta}_1(t)\leq0$, and $\dot{\eta}_2(t)\leq0$, the first two terms in both $\dot{A}(t)$ and $\dot{B}(t)$ are always negative. The sign of the last term, however, depends on the relative magnitudes of the decay parameters. Consequently, the characterization naturally separates into the following two cases.
\begin{itemize}
    \item If $\eta_1(t)>\eta_2(t)$, then the last term is also negative, making both $\dot{A}(t)$ and $\dot{B}(t)$ negative for all time. Hence, the resulting dynamics is BLP-Markovian. In this case, $\gamma_+(t)$ remains positive throughout the evolution, whereas $\gamma_-(t)$ and $\gamma_z(t)$ can become negative. Therefore, CP-divisible, P-divisible, and non-P-divisible dynamics can all be realized while remaining BLP-Markovian.

    \item If $\eta_1(t)<\eta_2(t)$, the last term becomes positive and can overcome the first two negative terms during part of the evolution. Consequently, $\dot{A}(t)$ and/or $\dot{B}(t)$ may become positive, giving rise to BLP non-Markovian dynamics. Even in this case, $\gamma_+(t)$ remains positive throughout the evolution, whereas the signs of $\gamma_-(t)$ and $\gamma_z(t)$ depend on the relative values of the decay parameters and their derivatives. Hence, all dynamical classes, including BLP-non-Markovian dynamics, can be realized.
    
\end{itemize}

\textbf{\textit{Conclusion:}}
Unlike the constant mixing probability case, a monotonically decreasing mixing probability can realize both BLP-Markovian and BLP-non-Markovian dynamics while both constituent channels remain CP-divisible. Consequently, all dynamical classes can be engineered. Throughout the evolution, $\gamma_+(t)$ remains positive, whereas the signs of $\gamma_-(t)$ and $\gamma_z(t)$ determine the nature of the resulting dynamics. 

\textbf{\textit{Case c.}} Finally, we allow the mixing probability to vary non-monotonically. Using Eqs.~(\ref{generalAB}) and (\ref{Generalrates}), the resulting dynamics is characterized as follows.

Since neither the sign nor the magnitude of $\dot{p}(t)$ is constrained, the characterization is no longer determined solely by the relative magnitudes of the decay parameters, but by the combined behavior of the decay parameters, the mixing probability, and their derivatives. Consequently, both $\dot{A}(t)$ and $\dot{B}(t)$ can become either positive or negative, allowing both BLP-Markovian and BLP-non-Markovian dynamics. Similarly, none of the three rates has a definite sign in general. Their signs depend on the interplay between the decay parameters, their derivatives, and the mixing probability. However, from the general rate expressions, it follows that the mixing-probability contribution enters $\gamma_+(t)$ and $\gamma_-(t)$ with opposite signs. Consequently, although either of the two rates may become negative, they can not become negative simultaneously.

\textbf{\textit{Conclusion:}}
With a non-monotonic probability, all dynamical classes can be realized while both constituent channels remain CP-divisible. None of the three rates has a definite sign in general, although $\gamma_+(t)$ and $\gamma_-(t)$ cannot become negative simultaneously.

As a final remark, irrespective of the divisibility or Markovian properties of the dynamics, the resulting channel becomes unital when the translation term vanishes, i.e., $X(t)=0$. This requires
\[p(t)=\frac{1-\eta_2(t)}{2-\eta_1(t)-\eta_2(t)},\]
provided the denominator is non-zero. Thus, unlike the equal-decay case, where unitality is achieved by the constant choice $p(t)=\frac{1}{2}$, here the required mixing probability generally depends on the decay parameters and is therefore time-dependent.

\textbf{Condition 2.}
Both constituent channels satisfy
\[\dot{\eta}_1(t)>0,\qquad \dot{\eta}_2(t)>0\]
during the same time intervals. Consequently, both constituent channels are non-CP-divisible. Since each constituent channel has only one non-zero rate, non-CP-divisibility is equivalent to non-P-divisibility. Furthermore, for these individual channels, P-divisibility is equivalent to BLP Markovianity. Thus, both constituent channels are BLP-non-Markovian.

We again distinguish the resulting dynamics according to the choice of mixing probability, with each case further divided into the corresponding subclasses.

\textbf{\textit{Case a.}} We first consider a constant mixing probability, i.e., $\dot{p}(t)=0$. Using Eqs.~(\ref{ConstantPAB}) and (\ref{rateConstantP}), the resulting dynamics is characterized as follows. 

Since both decay parameters increase simultaneously, $\dot{A}(t)$ and $\dot{B}(t)$ remain positive during these intervals, making the resulting dynamics BLP-non-Markovian. The same conditions also make both $\gamma_+(t)$ and $\gamma_-(t)$ negative simultaneously, irrespective of the relative values of the decay parameters. Consequently, the resulting dynamics is always non-P-divisible. Although $\gamma_z(t)$ does not affect these properties, its sign can be tuned through the relative values of $\eta_1(t)$, $\eta_2(t)$, and their derivatives.

\textbf{\textit{Case b.}} We now allow the mixing probability to decrease monotonically, i.e., $\dot{p}(t)\le0$ for all time. Using Eqs.~(\ref{generalAB}) and (\ref{Generalrates}), the resulting dynamics is characterized as follows.

The first two terms in both $\dot{A}(t)$ and $\dot{B}(t)$ are positive, whereas the contribution proportional to $\dot{p}(t)$ depends on the relative values of the decay parameters. Consequently, the characterization naturally separates into the following two cases:

\begin{itemize}
\item If $\eta_2(t)\geq\eta_1(t)$, the last term in both $\dot{A}(t)$ and $\dot{B}(t)$ is also positive. Hence, during the intervals where both decay parameters increase simultaneously, $\dot{A}(t)$ and $\dot{B}(t)$ remain positive, making the resulting dynamics BLP non-Markovian. This conclusion is independent of the values of the decay parameters, their derivatives, and the mixing probability.

\item If $\eta_2(t)<\eta_1(t)$, the last term becomes negative and competes with the positive contributions from the constituent channels. Consequently, the signs of $\dot{A}(t)$ and $\dot{B}(t)$ depend on the relative magnitudes of these competing terms. Therefore, both BLP-Markovian and BLP non-Markovian dynamics can be realized.
\end{itemize}

The rate $\gamma_-(t)$ becomes negative during the intervals where both decay parameters increase simultaneously, whereas the sign of $\gamma_+(t)$ depends on the relative values of the decay parameters, their derivatives, and the mixing probability. The sign of $\gamma_z(t)$ can again be tuned through these quantities. Consequently, the resulting dynamics is always non-P-divisible, irrespective of whether it is BLP-Markovian or BLP non-Markovian.

\textbf{\textit{Case c.}} Finally, we allow the mixing probability to vary non-monotonically. Using Eqs.~(\ref{generalAB}) and (\ref{Generalrates}), the resulting dynamics is characterized as follows.

Since the mixing probability is allowed to vary non-monotonically, $\dot{p}(t)$ is not restricted to remain non-positive throughout the evolution. Consequently, there exist intervals where $\dot{p}(t)>0$, although these need not coincide with the intervals during which both decay parameters increase simultaneously. The characterization therefore naturally separates into the following two cases:

\begin{itemize}
\item If $\dot{p}(t)>0$ during the intervals where both $\dot{\eta}_1(t)>0$ and $\dot{\eta}_2(t)>0$, the last term in both $\dot{A}(t)$ and $\dot{B}(t)$ is also positive. Hence, $\dot{A}(t)$ and $\dot{B}(t)$ remain positive during these intervals, making the resulting dynamics BLP non-Markovian. This conclusion is independent of the values of the decay parameters and their derivatives.

\item If $\dot{p}(t)<0$ during these intervals, the last term competes with the positive contributions from the constituent channels. Consequently, the signs of $\dot{A}(t)$ and $\dot{B}(t)$ depend on the relative values of the decay parameters, their derivatives, and the mixing probability. Therefore, both BLP-Markovian and BLP non-Markovian dynamics can be realized.
\end{itemize}

Among the rates, at least one of $\gamma_+(t)$ and $\gamma_-(t)$ necessarily becomes negative during the intervals where both decay parameters increase simultaneously, irrespective of the sign of $\dot{p}(t)$. The other rate, together with $\gamma_z(t)$, depends on the relative values of the decay parameters, their derivatives, and the mixing probability. Consequently, the resulting dynamics is non-P-divisible, irrespective of whether it is BLP-Markovian or BLP non-Markovian.

\textbf{Conclusion:}
Irrespective of the choice of the mixing probability, the resulting dynamics is always non-P-divisible since at least one of the rates $\gamma_+(t)$ and $\gamma_-(t)$ necessarily becomes negative during the intervals where both decay parameters increase simultaneously. The distinction between the three cases lies solely in the BLP behavior: while a constant mixing probability always produces BLP-non-Markovianity, relaxing the constraints on the mixing probability makes the BLP behavior increasingly dependent on the relative values of the decay parameters, their derivatives, and the mixing probability itself.

The above analysis assumes that the intervals over which both decay parameters increase coincide. When these intervals do not overlap, one decay parameter increases while the other decreases during the corresponding time intervals. In this situation, the fixed sign structure of $\dot{A}(t)$, $\dot{B}(t)$, and the rates is lost, and their behavior depends on the relative values of the decay parameters, their derivatives, and the mixing probability. A similar competing sign structure also arises when one constituent channel is CP-divisible while the other is non-CP-divisible. Since both situations admit the same type of parameter-dependent analysis, they are discussed together in the following condition.

\textbf{Condition 3.}
The Individual constituent channels satisfy
\[\dot{\eta}_1(t)\leq0, \qquad
\dot{\eta}_2(t)>0\]
throughout the evolution (or vice versa). Consequently, one constituent channel is CP-divisible, whereas the other is non-CP-divisible. Since each constituent channel has only one nonzero rate, the latter is simultaneously non-P-divisible and BLP-non-Markovian, whereas the former remains P-divisible and BLP-Markovian.

The following analysis also applies to the non-overlapping situation discussed after Condition~2. In both situations, one decay parameter increases while the other decreases, so that the fixed sign structure of $\dot{A}(t)$, $\dot{B}(t)$, and the rates is lost. Consequently, no universal analytical characterization can be made independent of the relative values of the decay parameters, their derivatives, and the mixing probability.

The two situations differ, however, in their temporal extent. In the non-overlapping case of Condition~2, this competing sign structure occurs only during the intervals where the increasing behavior of the two decay parameters does not coincide, whereas in the present condition it persists throughout the evolution. The resulting dynamics is therefore characterized through representative examples illustrating the different dynamical classes that can emerge.

\begin{figure}
    \centering
    \begin{subfigure}{0.48\textwidth}
        \includegraphics[scale=0.49]{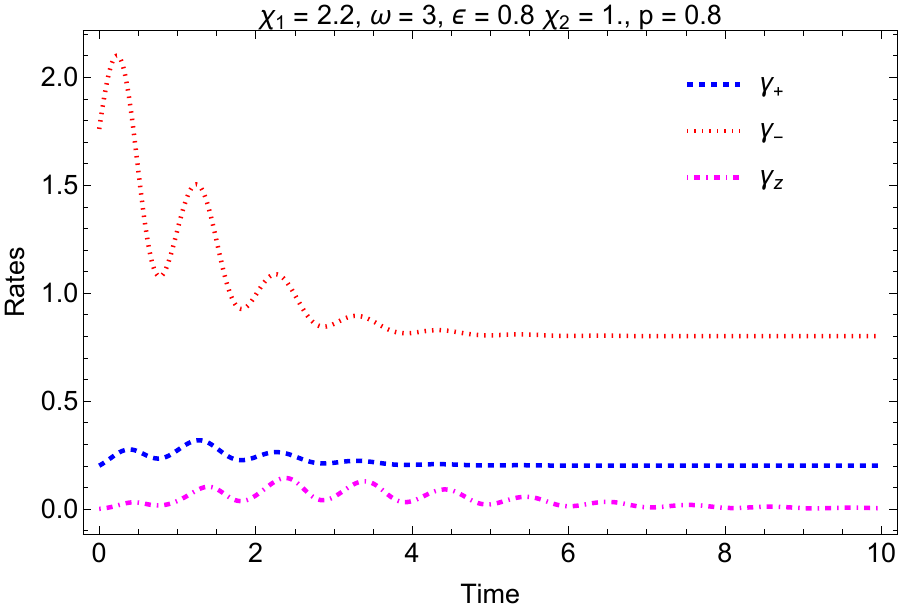} 
    \end{subfigure}
    \begin{subfigure}{0.48\textwidth}
        \includegraphics[scale=0.49]{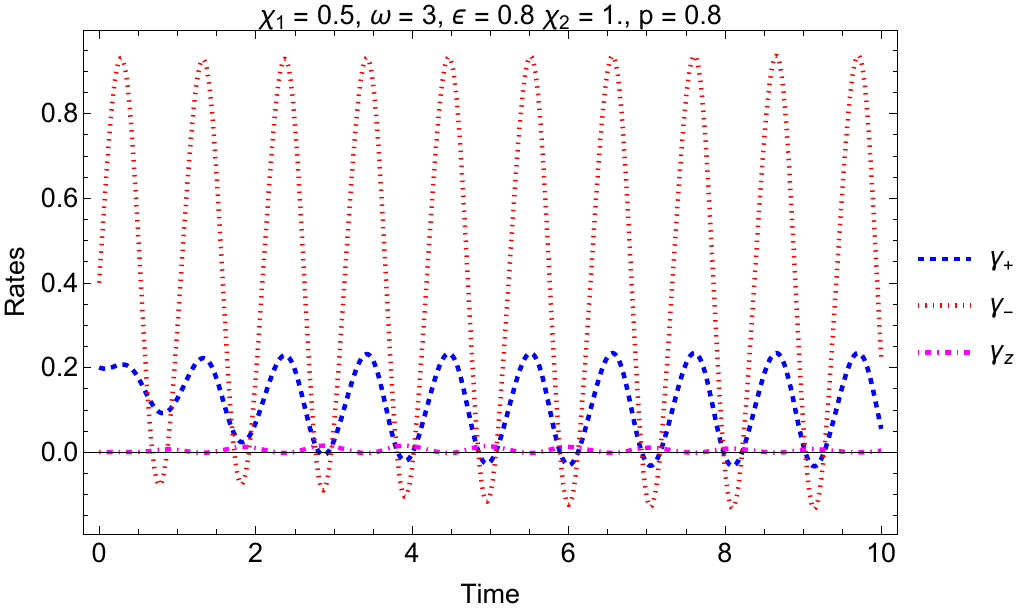} 
    \end{subfigure}\\
    \begin{subfigure}{0.48\textwidth}
        \includegraphics[scale=0.49]{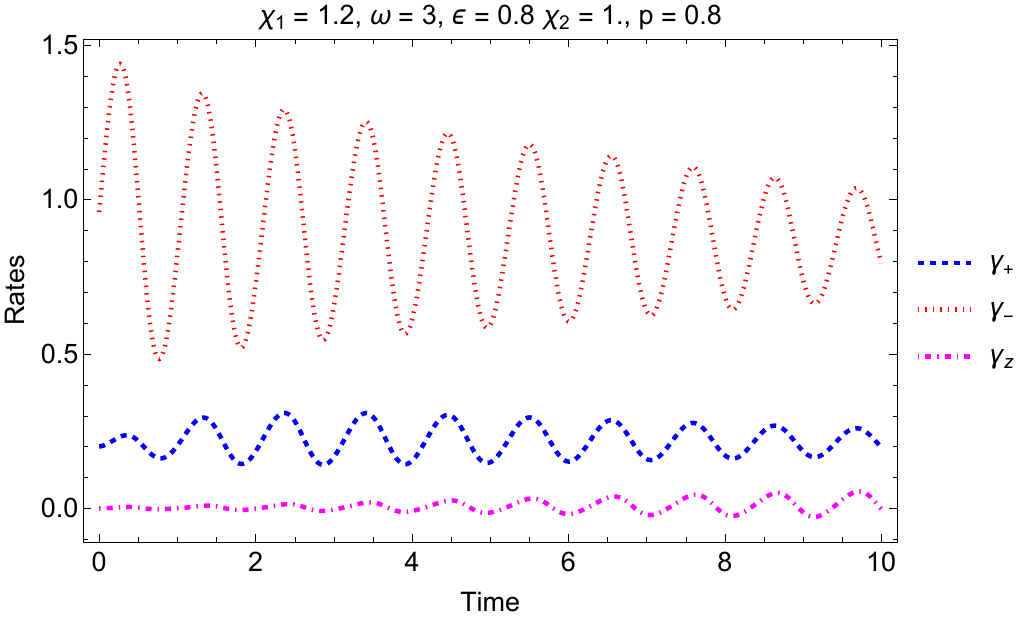}
    \end{subfigure}
    \begin{subfigure}{0.48\textwidth}
        \includegraphics[scale=0.49]{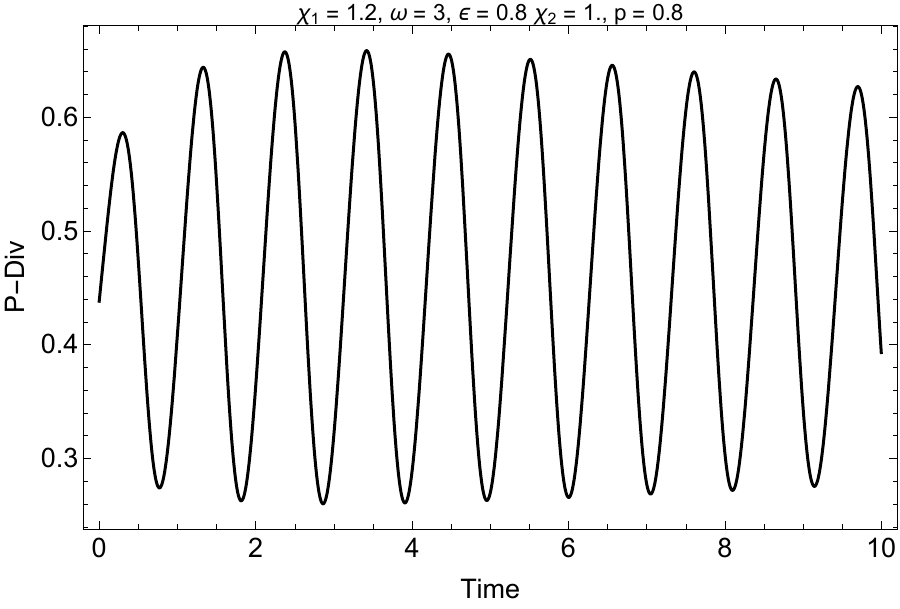}
    \end{subfigure}
    \begin{subfigure}{0.48\textwidth}
        \includegraphics[scale=0.49]{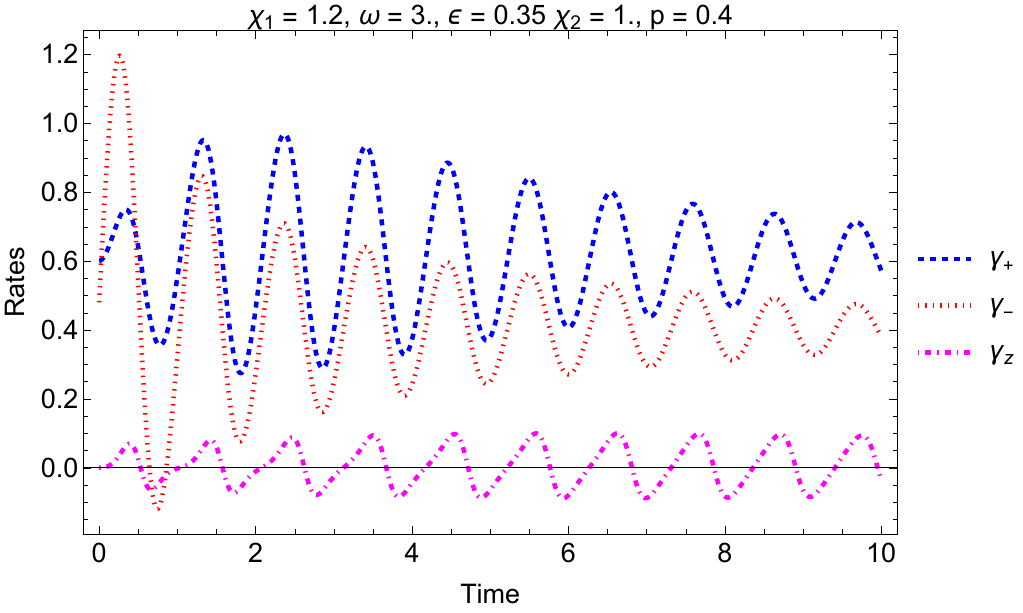}
    \end{subfigure}
    \begin{subfigure}{0.48\textwidth}
        \includegraphics[scale=0.49]{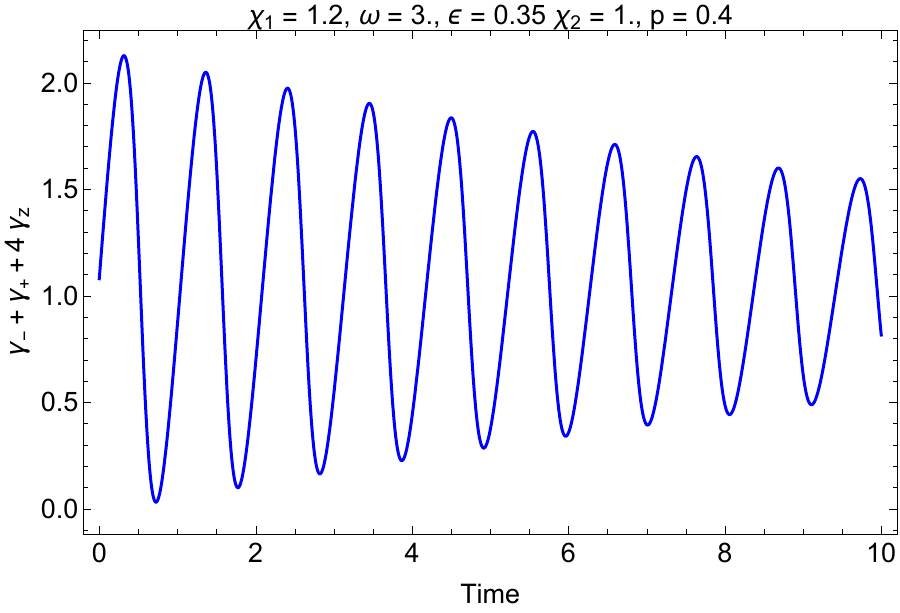}
    \end{subfigure}
    
    \caption{\justifying
Representative dynamical classes for the mixing of one CP-divisible and one non-P-divisible constituent channel with constant mixing probability $p$. The non-P-divisible channel has the oscillatory parameter $\eta_1(t)=(\epsilon+(1-\epsilon)\cos^2(\omega t))\exp[-\chi_1 t]$, while the CP-divisible channel has the monotonically decaying parameter $\eta_2(t)=\exp[-\chi_2 t]$. \textbf{Top left:} all three canonical rates remain positive, corresponding to CP-divisible dynamics. \textbf{Top right:} $\gamma_+(t)$ and $\gamma_-(t)$ become negative simultaneously, giving non-P-divisible and BLP non-Markovian dynamics. \textbf{Middle left:} $\gamma_z(t)$ becomes negative, indicating non-CP-divisible dynamics. \textbf{Middle right:} the P-divisibility condition $2\gamma_z(t)+\sqrt{\gamma_+(t)\gamma_-(t)}$ remains positive; together with the middle-left panel, this demonstrates P-divisible but non-CP-divisible dynamics. \textbf{Bottom left:} $\gamma_-(t)$ becomes negative, indicating non-P-divisible dynamics, while $\gamma_+(t)+\gamma_-(t)>0$ remains satisfied. \textbf{Bottom right:} the second BLP condition
$\gamma_+(t)+\gamma_-(t)+4\gamma_z(t)$ remains positive. Together with $\gamma_+(t)+\gamma_-(t)>0$, the two BLP conditions
establish BLP Markovianity for the bottom case. The different dynamical classes are obtained by varying the decay parameters of the constituent channels while keeping the mixing probability constant.}
    \label{CP_nonPMix}
\end{figure}

\textbf{Conclusion:}
Unlike the previous two conditions, no universal analytical characterization is possible in this competing-sign regime. The resulting dynamics is entirely determined by the relative values of the decay parameters, their derivatives, and the mixing probability. Consequently, all dynamical properties must be inferred from the complete expressions for the canonical rates and the corresponding BLP quantities, and representative examples are therefore required to illustrate the different dynamical classes that can emerge. A constant
mixing probability is already sufficient to realize the different dynamical classes; allowing the mixing probability to vary with time
does not introduce new dynamical classes, but provides additional control over their temporal structure. The different possible dynamics for constant probability is shown in Fig.~\ref{CP_nonPMix}. 

\textbf{Condition for non-invertibility:} The determinant of the map vanishes whenever either $A(t)=0$ or $B(t)=0$. Since $p(t)$, $\eta_1(t)$, and $\eta_2(t)$ are non-negative for all times, neither $A(t)$ nor $B(t)$ can vanish if both constituent channels remain invertible. Thus, mixing two invertible channels always results in an invertible dynamical map, irrespective of the mixing probability. 

If one or both constituent channels are non-invertible, the resulting dynamics can also become non-invertible under suitable conditions. If only one channel is non-invertible at a given instant, the mixing probability must assign unit weight to that channel at that instant. If both channels become non-invertible simultaneously, the resulting dynamics is non-invertible irrespective of the mixing probabilities. On the other hand, if the two channels become non-invertible at different times, the mixing probability of the non-invertible channel must be unity at the corresponding instant. Therefore, unlike the invertible case, the realization of non-invertible dynamics depends on both the non-invertible instants of the constituent channels and the mixing probability. For the non-GAD case, the condition $\eta_1(t)=0=\eta_2(t)$ coincides with the corresponding GAD condition, whereas the remaining two possibilities differ because of the additional degree of freedom introduced by the generalized amplitude damping channels.

Except for the special case in which all states evolve identically after the non-invertible instant, non-invertibility necessarily breaks divisibility. Furthermore, every non-invertible evolution is BLP non-Markovian. Indeed, at the non-invertible instant, at least two distinct states evolve to the same state, causing their trace distance to vanish. If these states subsequently evolve differently, their trace distance must increase from zero, implying information backflow according to the BLP criterion.
\begin{table}[t]
\centering
\caption{Bath engineering guide for realizing different dynamical classes through suitable choices of constituent channels and mixing probability. The intervals of both non-P-Div are assumed to be the same; otherwise, the results will be within the third category.}
\label{tab:engineering}
\renewcommand{\arraystretch}{1.3}

\begin{tabular}{|p{2.5cm}|p{2.2cm}|p{2.1cm}|p{2.1cm}|p{2.1cm}|p{2.1cm}|p{4.5cm}|}
\hline
\textbf{Constituent channels} &
\textbf{Mixing probability} &
\textbf{CP-Div. (BLP M.)} &
\textbf{P-Div but non-CP-Div. (BLP M.)} &
\textbf{Non-P-Div. (BLP M.)} &
\textbf{Non-P-Div. (BLP NM.)} &
\textbf{Characteristic rate behavior} \\
\hline

\multirow{3}{2.5cm}[-3.5ex]{\centering Both CP-Div} &
Constant &
Possible &
Possible &
Possible &
Not possible &
$\gamma_z$ can be tuned independently while $\gamma_\pm$ remain positive. \\
\cline{2-7}

&
Monotonically decreasing &
Possible &
Possible &
Possible &
Possible &
$\gamma_+>0$ but other rates are not fixed a priori. \\
\cline{2-7}

&
Non-monotonic &
Possible &
Possible &
Possible &
Possible &
No canonical rate sign is fixed a priori. But $\gamma_+$ and $\gamma_-$ cannot be negative simultaneously.  \\
\hline

\multirow{3}{2.5cm}[-3.5ex]{\centering Both non-P-Div} &
Constant &
Not possible &
Not possible &
Not possible &
Always &
Both $\gamma_+(t)$ and $\gamma_-(t)$ become negative simultaneously. \\
\cline{2-7}

&
Monotonically decreasing &
Not possible &
Not possible &
Possible &
Possible &
$\gamma_-(t)$ is negative whenever individual maps increase. \\
\cline{2-7}

&
Non-monotonic &
Not possible &
Not possible &
Possible &
Possible &
At least one of $\gamma_\pm(t)$ is always negative. \\
\hline

\multirow{3}{2.5cm}[-2ex]{\centering One CP-Div \& one non-P-Div}
&
Constant
&
\multirow{3}{2.1cm}{Possible}
&
\multirow{3}{2.1cm}{Possible}
&
\multirow{3}{2.1cm}{Possible}
&
\multirow{3}{2.1cm}{Possible}
&
\multirow{3}{4.5cm}{\centering
No rate behavior is fixed a priori; time-dependent mixing provides additional control over the temporal structure of the effective rates.
}
\\
\cline{2-2}

&
Monotonically decreasing
&
&
&
&
&
\\
\cline{2-2}

&
Non-monotonic
&
&
&
&
&
\\
\hline
\end{tabular}
\end{table}

\section{Error Mitigation}
Let us consider the amplitude damping (AD) channel as the unavoidable environmental noise acting on the system, whose dynamics cannot be controlled directly. Recent work~\cite{cpb_34_12_124703} has considered quantum error mitigation specifically for amplitude-damping noise using a Z-mixed expression of the channel. The objective is to reduce the deviation of the effective channel from the ideal noiseless identity channel, following the general motivation of quantum error mitigation~\cite{RevModPhys.95.045005}. Instead, in our approach, we propose to engineer the effective dynamics by mixing the AD channel with a controllable anti-AD channel together with an independently engineered mixing probability. To quantify the performance of this error-mitigation strategy, we consider the trace norm distance between the effective dynamical map and the identity channel,

\begin{equation}
    D(t)=||\mathcal{E}_t-\mathcal{E}_0||
\end{equation}
where $||A||=\rm Tr \sqrt{A^\dagger A}$ and $\mathcal{E}_0$ denotes the Identity channel, and $\mathcal{E}_t$ is the effective dynamical map expressed in the vectorized basis. Since this distance is evaluated directly on the dynamical maps, it provides a state-independent measure of the deviation from the ideal noiseless evolution.

\begin{figure}
    \centering
    \begin{subfigure}{0.45\textwidth}
        \includegraphics[scale=0.5]{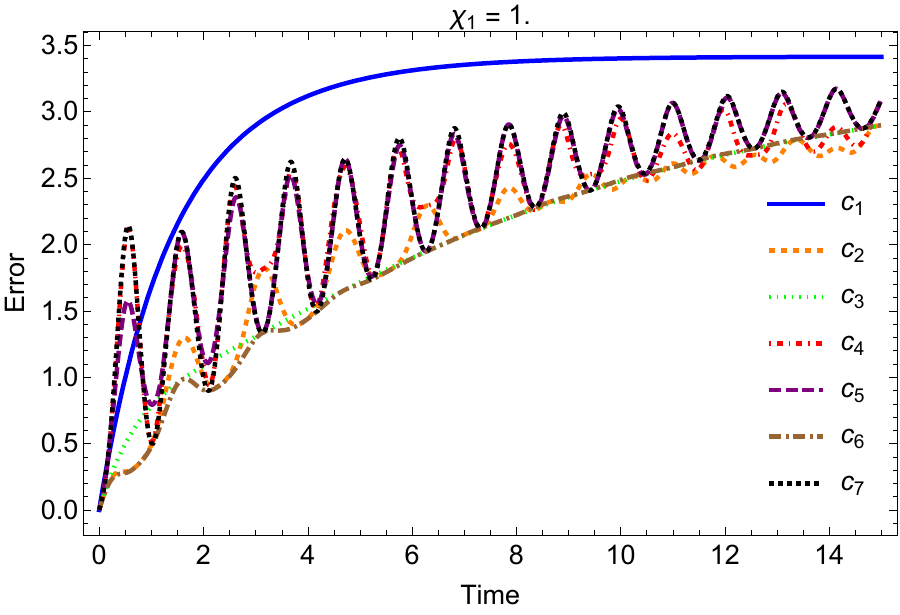} 
    \end{subfigure}
    \begin{subfigure}{0.45\textwidth}
        \includegraphics[scale=0.5]{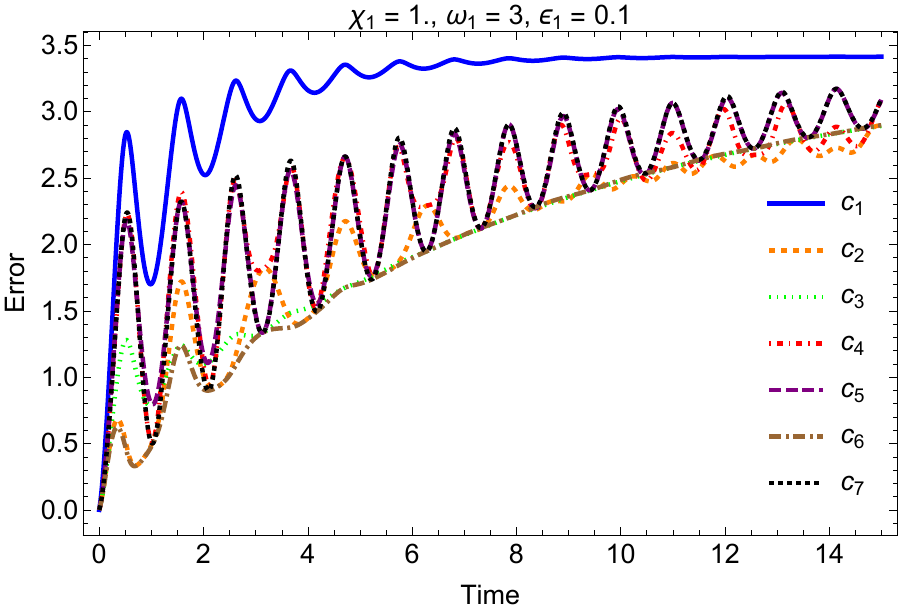} 
    \end{subfigure}
    
    \caption{\justifying
The error $D(t)$ is plotted as a function of time for different choices of the anti-amplitude-damping parameter $\eta_2(t)=(\epsilon_2+(1-\epsilon_2)\cos^2(\omega_2t))\exp[-\chi_2t]$ and the mixing probability $p(t)=p_0\exp[-\zeta t]\cos^2[\zeta_1 t]$. The curves $c_1, c_2....c_7$ correspond to Cases 1-7, respectively. \textbf{Left:} the AD channel is fixed with the monotonically decaying parameter $\eta_1(t)=\exp[-\chi_1t]$. \textbf{Right:} the AD channel is fixed with the oscillatory decaying parameter $\eta_1(t)=(\epsilon_1+(1-\epsilon_1)\cos^2(\omega_1t))\exp[-\chi_1t]$. Case 1 corresponds to the bare AD channel. Cases 2 and 4 have the same oscillatory mixing probability ($p_0=0.5$, $\zeta=0$, $\zeta_1=2$) but different anti-AD parameters ($\omega_2=3$, $\chi_2=0.2$, $\epsilon_2=1$ and $\epsilon_2=0.2$, respectively). Cases 3 and 5 similarly have the same decaying mixing probability ($p_0=0.5$, $\zeta=0.3$, $\zeta_1=0$) but different anti-AD parameters ($\omega_2=3$, $\chi_2=0.2$, $\epsilon_2=1$ and $\epsilon_2=0.2$, respectively), while Cases 6 and 7 have the same oscillatory decaying mixing probability ($p_0=0.5$, $\zeta=0.3$, $\zeta_1=2$) with different anti-AD parameters ($\omega_2=3$, $\chi_2=0.2$, $\epsilon_2=1$ and $\epsilon_2=0.2$, respectively). The different cases demonstrate the independent effects of engineering the mixing probability and the anti-AD channel on the deviation from the ideal noiseless evolution. Suitable choices can reduce the error, although some choices may temporarily increase it relative to the bare AD channel.}
    \label{ErrorMitigation}
\end{figure}

For the effective dynamical map, the trace norm can be evaluated analytically as 
\[D(t)=2(1-A(t))+\sqrt{X^2(t)+(1-B(t))^2}.\] 
The distance vanishes only when the effective channel coincides with the identity, i.e., when $A(t) = 1 = B(t)$ and $X(t) = 0$. These conditions can be satisfied in three different ways: (i) $p(t)=1$ with $\eta_1(t)=1$, (ii) $p(t)=0$ with $\eta_2(t)=1$, and (iii) $\eta_1(t) = 1 = \eta_2(t)$. These conditions require complete recurrence of at least one constituent channel, while the third possibility requires simultaneous recurrence of both constituent channels. 

The first possibility is generally unavailable since the AD channel is treated as the intrinsic environmental noise and cannot be manipulated. The second possibility would require a perfectly noiseless anti-AD channel together with a vanishing contribution from the AD channel, which is difficult to realize experimentally. The third possibility, however, provides a practical route whenever the controllable anti-AD channel exhibits complete recurrence. Since the anti-AD channel and the mixing probability are externally controllable, the timing and duration of the near-identity dynamics can be engineered to bring the effective dynamics arbitrarily close to the identity over a desired time interval. Such flexibility is absent when the constituent channels are constrained to have identical parameters, as in the standard GAD channel. 

Figure~\ref{ErrorMitigation} illustrates the temporal variation of the trace norm distance for different choices of the anti-AD channel and the mixing probability. Depending on the engineered parameters, the effective dynamics can exhibit a significant reduction in the error compared with the bare AD channel, although certain parameter choices may temporarily increase the deviation before the mitigation becomes effective.

\begin{figure}
    \centering
    \begin{subfigure}{0.45\textwidth}
        \includegraphics[scale=0.5]{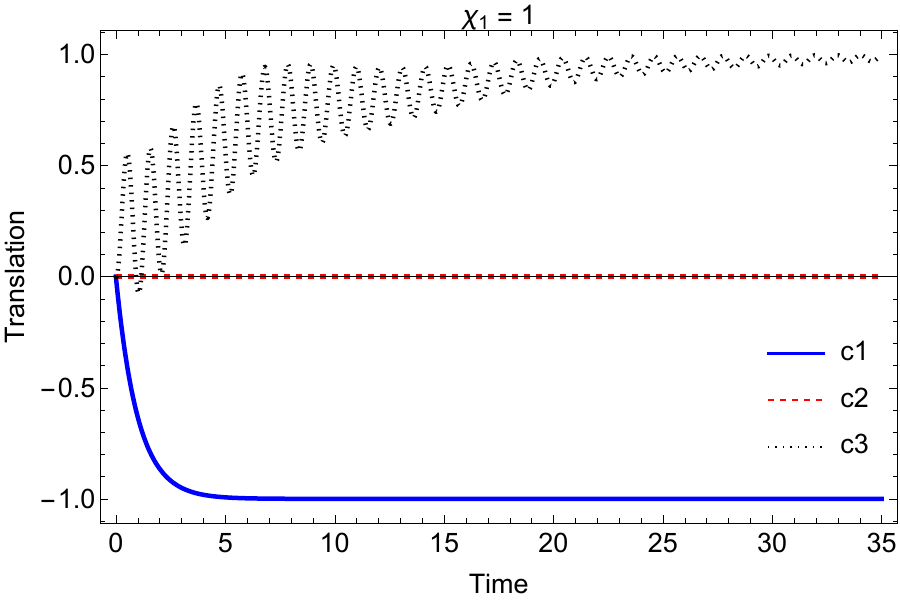} 
    \end{subfigure}
    \begin{subfigure}{0.45\textwidth}
        \includegraphics[scale=0.5]{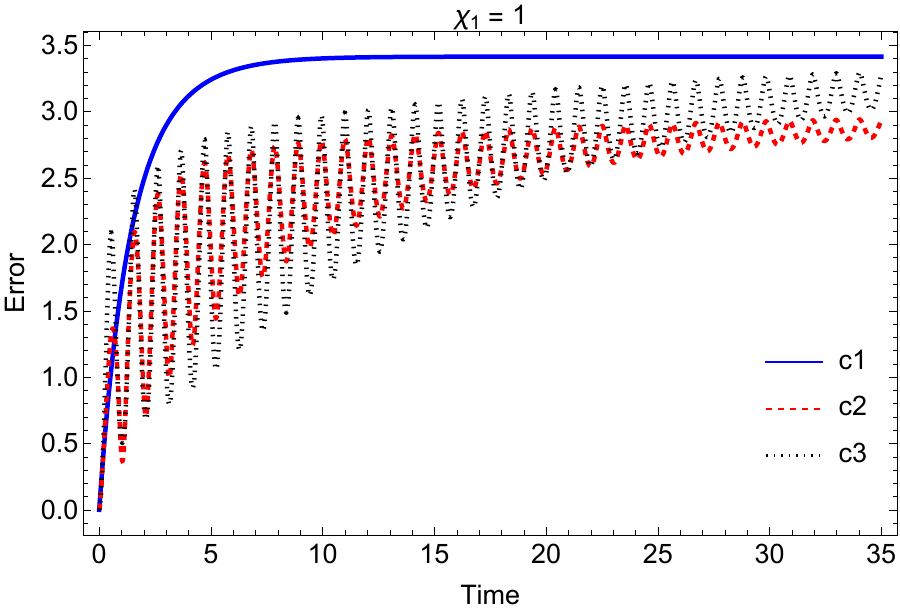} 
    \end{subfigure}\\
    \begin{subfigure}{0.45\textwidth}
        \includegraphics[scale=0.5]{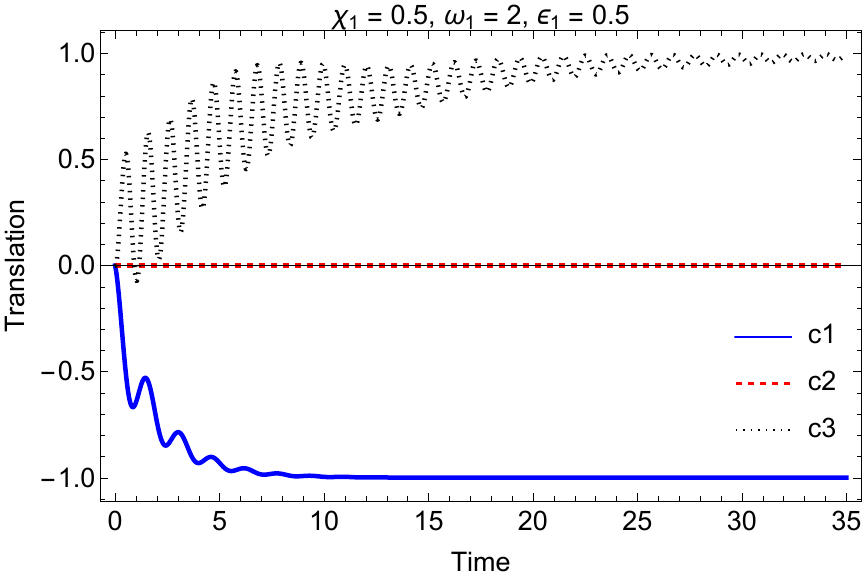}
    \end{subfigure}
    \begin{subfigure}{0.45\textwidth}
        \includegraphics[scale=0.5]{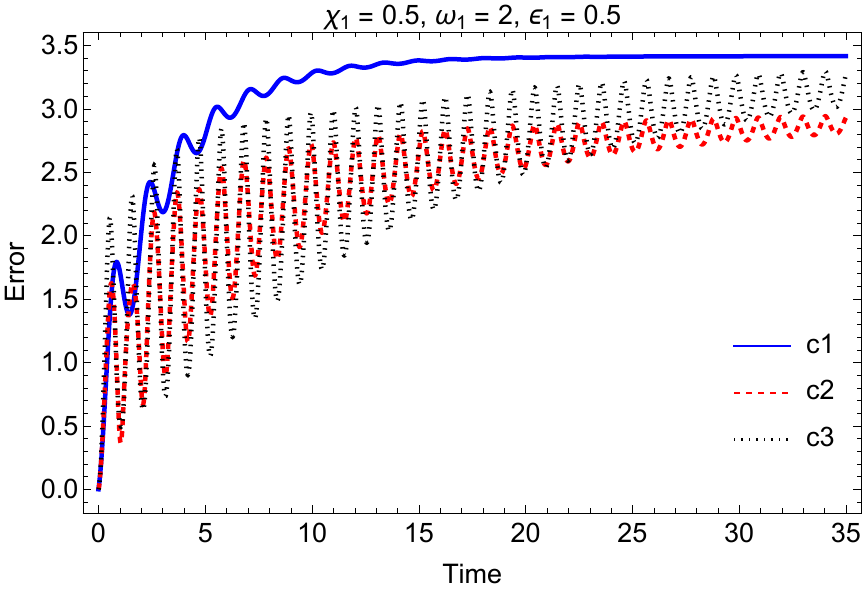}
    \end{subfigure}
    
    \caption{\justifying
Translation (left column) and error $D(t)$ (right column) are plotted as functions of time for two fixed AD channels and different choices of the anti-AD channel and mixing probability. The curves $c_1$, $c_2$, $c_3$ correspond to Cases 1, 2, and 3, respectively. \textbf{Top row:} monotonically decaying $\eta_1(t)=\exp[-\chi_1t]$. \textbf{Bottom row:} oscillatory decaying $\eta_1(t)=(\epsilon_1+(1-\epsilon_1)\cos^2(\omega_1t))\exp[-\chi_1t]$. \textbf{Case 1}, represented by the blue curves ($c_1$), correspond to the bare AD channel. \textbf{Case 2}, represented by the red curves ($c_2$), correspond to mixing with $\eta_2(t)=(\epsilon_2+(1-\epsilon_2)\cos^2(\omega_2t))\exp[-\chi_2t]$, with $\epsilon_2=0.1$, $\omega_2=3$, and $\chi_2=0.1$, where the mixing probability is chosen such that the effective channel is unital \Big($p(t)=\dfrac{1-\eta_2(t)}{2-\eta_1(t)-\eta_2(t)}$\Big). \textbf{Case 3}, represented by the black curves ($c_3$), use the same anti-AD channel as $c_2$, but with $p(t)=p_0\cos^2(\zeta_1 t)\exp[-\zeta t]$, where $p_0=0.3$, $\zeta_1=0.2$, and $\zeta=0.2$.}
    \label{Translation}
\end{figure}

If, instead, a unital channel close to the identity is desired, the two controllable functions, namely the mixing probability $p(t)$ and the anti-AD parameter $\eta_2(t)$, can be optimized simultaneously. Imposing the unitality condition yields 
\[p(t)=\frac{\eta_2(t)-1}{\eta_1(t)+\eta_2(t)-2}\]
which explicitly relates the required mixing probability to the given AD channel and the chosen anti-AD channel. This additional degree of freedom enables the translation term to be eliminated while the remaining anti-AD parameters can be optimized to minimize the overall deviation from the identity channel. This is shown in Figure.~\ref{Translation}.

\section{Discussion and Conclusion}

Amplitude damping is a ubiquitous source of decoherence in quantum systems and poses a significant obstacle to realizing quantum advantages. The corresponding anti-amplitude-damping process provides a natural framework for engineering effective open-system dynamics via channel mixing. When both channels have identical parameters, the resulting dynamics reduce to the generalized amplitude-damping channel, where the contraction of the Bloch sphere remains unchanged, and only the translation is modified. We first revisited this important special case by treating the channel parameter and mixing probability as independent quantities. This revealed that the GAD channel can exhibit non-P-divisible dynamics even when both constituent channels are individually CP-divisible, irrespective of whether the mixing probability is constant or time-dependent. In contrast, BLP non-Markovianity requires information backflow in the constituent dynamics. We also derived the condition for P-divisibility through an alternative approach based on a generalized theorem, which may be useful for analyzing other classes of quantum dynamical maps.

We then generalized the construction by allowing the amplitude-damping and anti-amplitude-damping channels to have completely independent parameters. This provides an independent route for engineering phase-covariant quantum channels by mixing two physically distinct constituent channels, with the mixing probability determining the degree of covariance. Since the resulting dynamics belong to the phase-covariant class, analytical conditions were derived to realize distinct dynamical behaviors characterized by CP-divisibility, P-divisibility, non-P-divisibility, and BLP non-Markovianity, as summarized in Table~\ref{tab:engineering}. These conditions were further translated into explicit constraints on the constituent channels and the mixing probability, providing a systematic framework for engineering the desired dynamical behavior. In contrast to the conventional GAD construction, in which the two constituent channels share the same channel parameters, the independent parametrization substantially enlarges the accessible dynamical space and allows the contraction and translation of the effective channel to be controlled through different combinations of constituent-channel parameters and mixing probability.

Finally, we demonstrated the practical utility of this construction for quantum error mitigation. Treating the amplitude-damping channel as uncontrollable environmental noise, the anti-amplitude-damping channel, together with the mixing probability, can be engineered to reduce the deviation of the effective dynamics from the ideal noiseless channel. The protocol can be applied to amplitude-damping noise with different temporal behaviors, including Markovian and BLP non-Markovian dynamics as well as different forms of divisibility. In addition to reducing overall error, the enlarged parameter space enables the realization of unital effective channels over a broader range of mixing probabilities than the conventional GAD construction, where the constituent channels share the same channel parameters. Even after imposing unitality, the remaining freedom in the anti-amplitude-damping channel can be exploited to further reduce the deviation from the ideal channel. Beyond error mitigation, the independent control of the translation term provides an additional route for quantum channel engineering, with potential applications in quantum communication, quantum sensing, and the modeling of open quantum dynamics in complex environments, including molecular and biological systems where energy exchange, decoherence, and environmental memory can play important roles.

\section*{Acknowledgments}

VP acknowledges the financial assistance of the Anusandhan National Research Foundation (ANRF) through grant CRG/2022/008345.	

\bibliography{main.bib}
\end{document}